# Emission Line Diagnostics for IMBHs in Dwarf Galaxies: Accounting for BH Seeding and ULX Excitation[*]

Chris T. Richardson,[1] Jordan Wels,[1] Kristen Garofali,[2] Julianna M. Levanti,[1] Vianney Lebouteiller,[3] Bret Lehmer,[4] Antara Basu-Zych,[5] Danielle Berg,[6] Jillian M. Bellovary,[7,8,9] John Chisholm,[6] Sheila J. Kannappan,[10] Erini Lambrides,[11] Mugdha S. Polimera,[12] Lise Ramambason,[13] Maxime Varese,[3] and Thomas Vivona[1]

[1]*Department of Physics & Astronomy, Elon University, 100 Campus Drive, Elon, NC 27244, USA*
[2]*William H. Miller III Department of Physics and Astronomy, Johns Hopkins University, Baltimore, MD 21218, USA*
[3]*Université Paris-Saclay, Université Paris Cité, CEA, CNRS, AIM, 91191, Gif-sur-Yvette, France*
[4]*Department of Physics, University of Arkansas, 226 Physics Building, 825 West Dickson Street, Fayetteville, AR 72701, USA*
[5]*NASA Goddard Flight Center, Code 662, Greenbelt, MD 20771, USA*
[6]*The University of Texas at Austin, 2515 Speedway, Stop C1400, Austin, TX 78712, USA*
[7]*Queensborough Community College, City University of New York, 222-05 56th Ave, Bayside, NY 11364, USA*
[8]*American Museum of Natural History, Central Park West at 79th Street, New York, NY 10024, USA*
[9]*Graduate Center, City University of New York, New York, NY 10016, USA*
[10]*Department of Physics & Astronomy, University of North Carolina, Phillips Hall CB 3255, Chapel Hill, NC 27599, USA*
[11]*NASA Goddard Space Flight Center, Code 662, Greenbelt, 20771, MD, USA*
[12]*Department of Physics & Astronomy, University of North Carolina, 141 Chapman Hall CB 3255, Chapel Hill, NC 27599, USA*
[13]*Institut fur Theoretische Astrophysik, Zentrum für Astronomie, Universität Heidelberg, Albert-Ueberle-Str. 2, D-69120 Heidelberg*

## ABSTRACT

Dwarf AGN serve as the ideal systems for identifying intermediate mass black holes (IMBHs) down to the most elusive regimes ($\sim 10^3 - 10^4 M_\odot$). However, the ubiquitously metal-poor nature of dwarf galaxies gives rise to ultraluminous X-ray sources (ULXs) that can mimic the spectral signatures of IMBH excitation. We present a novel photoionization model suite that simultaneously incorporates IMBHs and ULXs in a metal-poor, highly star-forming environment. We account for changes in $M_{BH}$ according to formation seeding channels and metallicity, and changes in ULX populations with post-starburst age and metallicity. We find that broadband X-rays and UV emission lines are insensitive to $M_{BH}$ and largely unable to distinguish between ULXs and IMBHs. Many optical diagnostic diagrams cannot correctly identify dwarf AGN. The notable exceptions include He II $\lambda4686$ and [O I] $\lambda6300$, for which we redefine typical demarcations to account for ULX contributions. Emission lines in the mid-IR show the most promise in separating stellar, ULX, IMBH, and shock excitation while presenting sensitivity to $M_{BH}$ and $f_{\rm AGN}$. Overall, our results expose the potential biases in identifying and characterizing dwarf AGN purely on strong line ratios and diagnostic diagrams rather than holistically evaluating the entire spectrum. As a proof of concept, we argue that recently discovered over-massive BHs in high-$z$ JWST AGN might not represent the overall BH population, with many galaxies in these samples potentially being falsely classified as purely star-forming.

*Keywords:* galaxies – dwarf, evolution – active galactic nuclei – intermediate-mass black holes

Corresponding author: Chris Richardson
crichardson17@elon.edu

[*] 

## 1. INTRODUCTION

A census of intermediate-mass black holes (IMBHs) spanning $\sim 10^2 - 10^5\ M_\odot$ is essential for a holistic understanding of galaxy - black hole (BH) co-evolution. BH seeds in the IMBH regime at high-$z$, which gave rise to supermassive BHs in the nuclear centers of giant galaxies, still elude current observational facilities (Greene et al. 2020). However, IMBHs that remain relatively un-evolved throughout cosmic time provide a unique test bed for various seeding mechanisms (Mezcua 2017), including Population III stars (light seeds), runaway stellar mergers, and direct collapse BHs (heavy seeds).

Based on the $M_{BH} - M_\star$ relation (Reines & Volonteri 2015), dwarf galaxies ($M_\star \leq 10^{9.5} M_\odot$) are likely to host nuclear BHs in the mass range of IMBHs. Indeed, models and extrapolated observations suggest that the BH occupation fraction in dwarf galaxies can range anywhere from 0.85 to unity for dwarfs with $M_\star \sim 10^9 M_\odot$ (Volonteri et al. 2008; Miller et al. 2015; Ricarte & Natarajan 2018; Greene et al. 2020; Cho & Woo 2024). However, high spatial resolution is required for dynamically identifying IMBHs at even modest distances $\sim 10$ Mpc (Nguyen et al. 2018). This limitation makes active galactic nuclei (AGN) an attractive option for finding IMBHs in dwarfs and establishing a limit on the BH occupation fraction in this regime.

While line broadening from the broad line region and accretion disk is the gold standard for powerful AGN detection, it becomes difficult to disentangle broadening mechanisms for the most elusive BHs with $\lesssim 10^5 M_\odot$. As a result, indirect detection through spectral signatures from ionized gas surrounding the AGN provides the most promising means for finding large samples of active IMBHs in dwarfs.

Both theory and observation have shown that emission line diagnostics for giant AGN often fail for dwarf galaxies hosting AGN (dwarf AGN; Reines et al. 2020; Polimera et al. 2022; Richardson et al. 2022; Dors et al. 2024; Mezcua & Domínguez Sánchez 2024). In particular, the BPT diagram (Baldwin et al. 1981) fails for metal-poor ($\lesssim 0.4$ Z$_\odot$), star-forming galaxies like dwarfs due to the metallicity sensitivity of [N II] $\lambda$6584 and due to their hard ionizing spectra at low metallicities causing high [O III]/H$\beta$. Likewise, emission lines associated with carbon species show sensitivity to the adopted abundance patterns on account of complex prescriptions for pseudo-secondary nucleosynthesis (Henry et al. 2000; Kobayashi et al. 2011; Berg et al. 2019).

In response, recent studies have focused on identifying emission line diagnostics that successfully identify AGN in metal-poor, highly star-forming conditions. In the optical, [Ne V] $\lambda$3427, He II $\lambda$4686, and [O I] $\lambda$6300 have been successful in separating dwarf AGN from star-forming galaxies when the fraction of the ionizing photons associated with the AGN ($f_{AGN}$) is $\geq 16\%$ (Sartori et al. 2015; Reines et al. 2020; Mezcua & Domínguez Sánchez 2020; Polimera et al. 2022; Feltre et al. 2023; Chisholm et al. 2024). In the mid-IR, [Ar II] 6.98$\mu$m, [Ne V] 14.3$\mu$m, and [O IV] 25.9$\mu$m show promise in separating dwarf AGN from star-forming galaxies down to the lowest $f_{AGN}$ (Richardson et al. 2022; Garofali et al. 2024). However, the simulations used to justify dwarf AGN diagnostics usually focus on one particular wavelength band rather than comprehensively covering emission from the IR to X-ray. Emission lines frequently seen, although not often used, in identifying AGN have been gaining popularity in the JWST era (e.g., [Ne III] $\lambda$3869 and [O II] $\lambda$3727) but lack the detailed modeling compared to more common lines like [O III] $\lambda$5007.

Photoionization models of dwarf AGN often make assumptions about the properties of the excitation sources in a dwarf galaxy with varying degrees of validity (e.g., Richardson et al. 2022, Satyapal et al. 2021, Cleri et al. 2023). For example, stellar population synthesis (SPS) models where stars evolve without a companion star is one such assumption (e.g., Starburst 99; Leitherer et al. 1999). More realistic SPS models that include binary evolution show that young Wolf-Rayet (WR) stars can produce ionizing radiation much harder than those resulting from models where stars evolve in isolation. Indeed, the WRs and stripped stars (Götberg et al. 2018) originating from binarity generate ionizing continua on timescales <25 Myr that can mimic low excitation AGN, particularly at low metallicities (Xiao et al. 2018; Lecroq et al. 2024). The stochastic, bursty star formation histories of dwarfs and their metal-poor gas emphasize the need to include SPS models that account for binary systems in photoionization models.

Simulations of dwarf AGN have also neglected the presence of non-thermal excitation sources other than AGN. At low metallicities ($\lesssim$ 0.4 Z$_\odot$), X-ray binary (XRB) populations become more prevalent due to weaker stellar winds (Linden et al. 2010; Mapelli et al. 2010; Wiktorowicz et al. 2017). In a luminous XRB, an accreting compact object generates thermal disk emission that peaks in the X-rays and often contains a non-thermal component due to winds and disk coronae. Coupled with stellar sources, such a radiation field can result in emission line signatures similar to AGN (Simmonds et al. 2021); however, the predictions from such models depend strongly on the shape of the intrinsic XRB spectral energy distribution (SED) and the scaling relations used to couple the XRB with the stellar population (e.g., $L_X$/SFR) and $Z$, both of which are empirically uncertain (Brorby et al. 2016; Lehmer et al. 2021).

Photoionization models often incorrectly assume that AGN SED does not evolve with cosmic time (e.g., Dors et al. 2024). In reality, radiation from the accretion disk in IMBHs shifts to the EUV and soft X-rays as $M_{\rm BH}$ decreases (Peterson 1997), which is expected in low-mass, metal-poor dwarf galaxies. The resulting SEDs vary greatly from those found in local giant AGN, and thus, emission line diagnostics calibrated to find those AGN can possibly fail in the dwarf regime (Cann et al. 2019). More generally, many emission line predictions are more sensitive to varying $M_{\rm BH}$ over several dex rather than other components to the SED, as usually performed for giant AGN (e.g., $\alpha_{\rm X}$, $\Gamma$; Thomas et al. 2018). The exact $M_{\rm BH}$ expected within dwarf galaxies of a given stellar mass is uncertain and dependent on the seeding mechanism, merger history, and accretion history of the IMBH.

Finally, the ubiquitously star-forming nature of dwarfs (Geha et al. 2012) suggests that emission line predictions for dwarf AGN should ultimately focus on AGN contributing a relatively low fraction to the overall ionizing continuum. Consequently, photoionization models including only AGN excitation will likely over-predict the emission from high-ionization lines.

In summary, a photoionization model suite is desperately needed that simultaneously includes WR, XRB, and IMBH excitation in a self-consistent way for metal-poor, highly star-forming (low $f_{\rm AGN}$) dwarf galaxies over a wide range of stellar masses. Robustly analyzing emission line predictions spanning the IR to the UV is necessary to thoroughly understand the biases involved in each wavelength regime as present-generation observatories (e.g., JWST) and next-generation telescopes (e.g., ELT, PRIMA, HWO) provide increased sensitivity to nebular emission from galaxies across cosmic time. In this paper, we present photoionization models to achieve this goal. We cohesively couple multiple prescriptions for different excitation sources while accounting for factors such as post-starburst age and IMBH seeding mechanisms, which can influence the contributions of each source. By making multi-wavelength emission line predictions to find active IMBHs in $z \lesssim 0$ dwarf galaxies, we can assess the most reliable and unreliable IMBH diagnostics available for multiple observatories.

In Section 2, we describe our methodology for scaling the $M_{\rm BH}$ with metallicity by using the $M_\star - Z$ and $M_{\rm BH} - M_\star$ scaling relations. In Section 3, we outline the setup of our photoionization models with a particular emphasis on understanding the nature of the incident SEDs resulting from simultaneous ultraluminous X-ray sources (ULXs) and IMBH activity. In Section 4, we assess the variation of common emission line ratios (IR-UV) due to the scaling relationships present in Section 2. In Section 5, we assess the sensitivity of emission lines to increasing AGN contributions while contextualizing their brightness relative to H$\alpha$. In Section 6, we determine the level of AGN contribution required to the washout signature of a ULX for both broadband X-ray (2–10 keV) observations and emission lines (UV-IR). In Section 7, we determine the physical conditions where emission line ratio diagrams succeed in separating IMBHs from other excitation sources and the conditions where false positive detection of IMBHs can occur due to ULXs. In Section 8, we highlight the diagnostic diagrams most promising for identifying dwarf AGN based on the results in previous sections while creating demarcations to indicate possible ULX contributions. In Section 9, we connect how our results can be used in future observations of $z \sim 0$ dwarf galaxies and contextualize the findings of recent JWST observations at high-$z$. Finally, in Section 10, we review our key results. We make our photoionization models freely available to the community[1].

## 2. CONNECTING BLACK HOLE MASS WITH GAS PHASE METALLICITY

In this section, we leverage galaxy scaling mass-metallicity and $M_\star$ vs. $M_{\rm BH}$ relations to tie gas-phase metallicity to BH mass. Both $Z$ and $M_{\rm BH}$ affect the overall ionizing SED since the stellar and ULX populations change as a function of gas-phase metallicity, and the $M_{\rm BH}$ changes the accretion disk temperature.

### 2.1. Mass-Metallicity Relations

The stellar mass ($M_\star$) and gas phase metallicity relation (hereafter, MZR) in star-forming galaxies has been extensively studied (Tremonti et al. 2004; Kewley & Ellison 2008; Berg et al. 2012; Curti et al. 2020). One reason for the positive correlation between $M_\star$ and Z stems from more massive galaxies having larger gravitational potential wells that inhibit metal-rich supernovae ejecta from escaping to the intergalactic medium (Mac Low & Ferrara 1999).

The MZR illustrates that dwarf galaxies (log $M_\star \leq$ 9.5) have low metallicities, and therefore the metallicity sensitive [N II] $\lambda 6584$ used in the BPT diagram results in dwarf AGN classified as star-forming galaxies. Understanding the MZR is paramount to connecting photoionization model predictions for emission lines like [N II] $\lambda 6584$ to global galaxy properties like $M_\star$. The MZR depends on star formation rate (SFR; Mannucci et al. 2010) and the redshift of the sample (Torrey et al. 2019; Langeroodi et al. 2023). For simplicity, we restrict our analysis by considering MZRs derived from $z \sim 0-1$ galaxies without corrections due to SFR.

---

[1] https://github.com/crichardson17

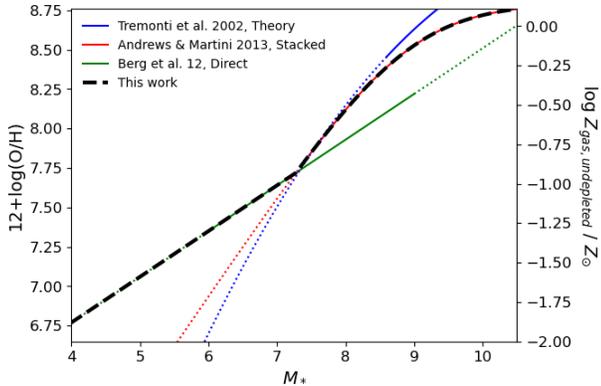

**Figure 1.** The mass-metallicity relation for local galaxies using three different techniques. The dotted lines show extrapolated regions for each technique. The MZR adopted in this paper (black dashed line) results from merging the predictions of two different regimes similar to Indahl et al. (2021).

Most methods of determining the MZR are based upon two approaches: (1) photoionization models are fit to spectroscopic observations to infer $Z$, which we will call the "theoretical method"; (2) electron temperatures ($T_e$) are directly measured from observed $T_e$-sensitive collisionally-excited lines (e.g., [O III] $\lambda 4363/\lambda 5007$), or recombination lines, and used to calculate gas-phase abundances directly, which we call the "direct method". The two methods produce very different results and suffer from different forms of bias.

For galaxies at solar metallicity and above, the theoretical method overpredicts the metallicity of supergiants in the Milky Way and Large Magellanic Cloud (Maiolino & Mannucci 2019). In contrast, the direct method underpredicts the metallicity of more metal-rich galaxies since the auroral lines needed for analysis are biased to higher temperatures (lower metallicities). Similarly, MZRs determined from the direct method rarely extend up to solar metallicity (Lebouteiller et al., accepted). A compromise is stacking galaxies into stellar mass bins to obtain the S/N required for the direct method rather than initially selecting galaxies with adequate S/N, however strong radial gradients in these more massive systems make it difficult to compare derived metallicities across a wide range of galaxy masses. Figure 1 shows differences in the theoretical, direct, and stacked methods of determining the MZR according to Tremonti et al. (2004), Berg et al. (2012), and Andrews & Martini (2013), respectively. For this paper, we select the calibration from Andrews & Martini (2013) for metallicities closer to solar since it agrees with the observations of more metal-rich galaxies.

The gas-phase metallicities used have been depleted due to the formation of dust grains. We account for this depletion by adding 0.11 dex to the oxygen abundance before scaling to our adopted solar value of $12+\log(O/H) = 8.76$. The resulting metallicity (hereafter referred to as "$Z$") is akin to the metallicity of the stars formed out of the gas in which they are exciting. Figure 1 shows that MZRs using the theoretical method (blue line), or stacked spectra (red line), rarely extend to the lower metallicity regime ($\lesssim 0.1$ $Z_\odot$). In metal-poor regions, stars have higher effective temperatures, and coolants have been removed from the gas, which translates to increased electron temperatures; thus, the direct method (green line) becomes much easier to apply and preferable at the lowest metallicities. Therefore, we select the calibration from Berg et al. (2012) in this regime.

Our overall MZR in Figure 1 (black dashed line) comes from merging two different calibrations across a wide range of subsolar metallicities. Such an approach is consistent to similar relations presented in Indahl et al. (2021) and valid up to $z \sim 1$ in both the higher metallicity (Torrey et al. 2019) and lower metallicity (Pharo et al. 2023) regimes.

All of the aforementioned relations, and therefore the one we adopt in this paper, only pertain to star-formation dominated galaxies. The effect of AGN on the MZR across a wide range of stellar masses is unclear. At stellar masses $M_\star \geq 10^{10} M_\odot$, AGN show a 0.1 dex increase in metallicity from the local MZR determined using the theoretical method (Thomas et al. 2018; Li et al. 2024). The stellar masses to which this result applies translate to the upper end of metallicities considered in this paper and, therefore, are largely inapplicable to our derived MZR scaling.

### 2.2. $M_{\rm BH}$ vs. $M_\star$ Relations

The correlation between nuclear $M_{\rm BH}$ and galaxy stellar mass has profound implications for BH-galaxy co-evolution. Incorporating this relation into photoionization models means accounting for the evolution of $M_{\rm BH}$ as galaxies evolve, i.e., less massive galaxies have less massive nuclear BHs. Calibrating the relation for AGN depends upon the redshift of the sample and the galaxy mass regime. The local $M_{\rm BH}$ vs. $M_\star$ relation shows a power law scaling down to $\sim 10^9 M_\odot$ (Reines & Volonteri 2015; Greene et al. 2020) and holds up to $z \sim 2.5$ (Suh et al. 2020). Recent work has shown that overmassive BHs relative to the local relation exist at $z \sim 1-2$ (Mezcua et al. 2024), although they are likely outliers to the overall distribution.

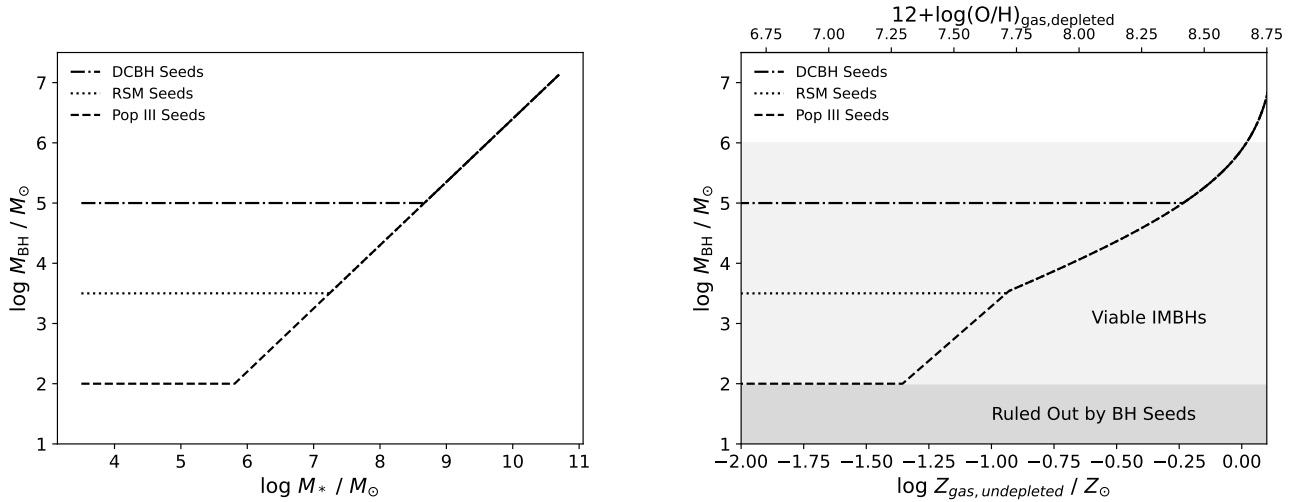

**Figure 2.** The $M_{\rm BH}$–$M_\star$ (left) and $M_{\rm BH}$–$Z$ (right) relations assumed in this paper. The seeding channels provide different evolutionary paths for IMBHs according to Population III stars (light seeds), runaway stellar mergers (intermediate seeds), and direct collapse black holes (heavy seeds). The $M_{\rm BH}$–$Z$ relation provides a convenient way to link the evolution of black hole masses with galaxy environment in photoionization simulations.

At high-$z$ ($4 < z < 11$), however, NIRSpec observations with JWST regularly show overmassive black holes in dwarf galaxies relative to the local relation approaching $M_{\rm BH} \approx M_\star$ (Harikane et al. 2023; Maiolino et al. 2023; Greene et al. 2024). There is much debate about whether the galaxies hosting these anomalously massive BHs are a representative sample of the entire AGN population at high-$z$ or subject to selection bias (Lupi et al. 2024).

No robust observational constraints on the $M_{\rm BH}$ vs. $M_\star$ relation exist for $M_{\rm BH} < 10^{4.5} M_\odot$ and finding IMBHs in this mass range is one of the main motivations for this paper. As such, simulations of IMBH - galaxy co-evolution must provide insight into the shape of the $M_{\rm BH}$ vs. $M_\star$ relation down to the lowest IMBH masses. Unfortunately, cosmological simulations (e.g., Illustris: Sijacki et al. 2015, EAGLE: McAlpine et al. 2016, SIMBA: Davé et al. 2019) provide different $M_{\rm BH}$–$M_\star$ scalings due to numerical choices, accretion efficiency, and seeding channels (Haidar et al. 2022). Additionally, these large-scale cosmological simulations rarely initialize (seed) their black holes at the masses necessary to compare to low stellar mass local dwarf galaxies. One workaround involves including sub-grid physics specifically to trace the growth of the lowest mass BH seeds. In these simulations, once IMBHs are seeded, they show little growth over large periods of time before dramatically increasing mass and converging upon local relations at $z < 1$ (Beckmann et al. 2023; Bhowmick et al. 2024).

Given the uncertainty in the $M_{\rm BH}$ vs. $M_\star$ relation at high-$z$, we use the relations as they apply to galaxies at $z < 1$, which have been more robustly studied. Following the results of Beckmann et al. (2023) and Bhowmick et al. (2024), we adopt a $M_{\rm BH}$ vs. $M_\star$ relation that follows the Reines & Volonteri (2015) until the mass threshold for various seeding channels is reached. We assume that Population III stars create BH seeds with $M_{\rm BH} = 10^2 M_\odot$ (i.e., "light seeds"), runaway stellar mergers (RSM) create BH seeds with $M_{\rm BH} = 10^{3.5} M_\odot$, and direct collapse black holes (DCBH) create BH seeds with $M_{\rm BH} = 10^5 M_\odot$ (i.e., "heavy seeds"; Greene et al. 2020). Figure 2 (right panel) displays the resulting $M_{\rm BH}$ vs. $M_\star$ relations for each seeding channel in conjunction with local relation, valid for $z < 1$ galaxies.

### 2.3. $M_{\rm BH}$ vs. $Z$ Relation

The stellar mass component to the MZR from Figure 1 and the $M_{\rm BH}$ vs. $M_\star$ relations in Figure 2 enables a relation between metallicity and IMBH mass. The resulting $M_{\rm BH}$ vs. $Z$ relations are shown in the right panel of Figure 2. The Population III (light) seeding channel shows that the local $M_{\rm BH}$–$M_\star$ relation *must* deviate at $\log Z/Z_\odot \approx -1.4$ or else the predicted BH masses become too low to reconcile with BH seeding mechanisms.

For light seeds, the exact expression is given as,

$$M_{\rm BH} = \begin{cases} 5.25 - 1.64 \log(10^{0.15-\tilde{Z}} - 1) & \tilde{Z} \geq \tilde{Z}_1 \\ 3.62\,\tilde{Z} + 6.91 & \tilde{Z}_2 \leq \tilde{Z} < \tilde{Z}_1 \\ 2.0 & \tilde{Z} < \tilde{Z}_2 \end{cases} \quad (1)$$

where $\tilde{Z} = \log Z/Z_\odot$, $\tilde{Z}_1 = -0.93$, and $\tilde{Z}_2 = -1.36$. The RSM and DCBH seeding channels deviate from the light seeds at $\tilde{Z} = -0.95$ and $\tilde{Z} = -0.24$, respectively. Overall, this $M_{\rm BH}$ vs. $Z$ relation creates a link between the gas, stars, and nuclear BHs in galaxies, thereby enabling substantial improvement in photoionization modeling designed to provide insights about detecting elusive IMBHs in dwarfs. While there is indeed observational scatter in $M_{\rm BH}$ vs. $Z$, especially at low metallicities, the limiting cases are valid: low mass BHs reside in low metallicity galaxies and high mass BHs reside in high metallicity galaxies. Our work establishes a first-order relationship to investigate how much this affects the emission line predictions for dwarf AGN.

## 3. PHOTOIONIZATION MODELS

### 3.1. Radiation Fields

In this section, we construct representative incident SED templates for stellar ionization sources that self-consistently include ULXs and have varying levels of AGN activity to eventually explore how stellar, ULX, and AGN contributions impact emission-line ratio diagnostics.

#### 3.1.1. Excitation Sources

**Stars** - The Binary Population and Spectral Synthesis (BPASS) models are used to understand the evolution of stellar populations, and their respective SEDs, simulating both single and binary star systems across a range of stellar masses and metallicities (Eldridge et al. 2017). We use BPASS to model stellar evolution as simple stellar populations (SSPs), which are stars formed simultaneously and with the same metallicity but can evolve differently, especially in binary systems. BPASS accounts for three stellar evolutionary paths for binary systems. First, binary star ionizing radiation can regenerate at the end of stellar lifetimes due to mass transfer, allowing old stars to disguise as much younger stars. Another evolution path that regenerates radiation includes the effect of star mergers, again allowing the end of a stellar lifetime to rejuvenate under the disguise of a young star (Schneider et al. 2016). Both evolutionary paths extend the lifetimes of high mass stars into ages very unexpected of high mass stars. The third process is envelope removal from Roche lobe overflow (Dray & Tout 2007), resulting in low-luminosity, high metallicity, WR stars, stripped stars, and helium stars at older ages than expected. This leaves a thin hydrogen layer that requires extreme solar winds to be removed (Smith 2014; Trani et al. 2014). A primary effect of including binary systems and their different evolutionary path is that the ionizing flux of the stellar spectra is increased by 50%–60% at low metallicities (D'Agostino et al. 2019).

In reality, the star formation histories (SFHs) of dwarf galaxies are rather bursty, undergoing multiple episodes of star formation, which are not accurately modeled by an SSP or continuous star formation (Domínguez et al. 2015). Since a primary focus of this paper is identifying models that can mimic AGN excitation, we choose to model the stellar component as an SSP where the effects of the WR phase are more prominent (Xiao et al. 2018).

The BPASS models are a function of the instantaneous burst age (i.e., post-starburst age) and $Z$, making them time and metallicity-dependent. Using BPASS version 2.2 (Eldridge et al. 2017), we selected the initial mass function (IMF) with a cutoff at 100 $M_\odot$, a low-mass (0.1–0.5 $M_\odot$) IMF slope of $-1.30$, and a high-mass (0.5–100 $M_\odot$) IMF slope of $-2.35$ (Kroupa 2002; Fragos et al. 2013). Our models are characterized by specific ages of 1, 2, 3.16, 3.98, 5.01, 7.94, 10, 15.8, 20, 25.1 Myr, and metallicities $\log Z/Z_\odot = -2, -1.3, -1, -0.7, -0.4, 0$ where $\log Z/Z_\odot = 0$ is considered $Z_\odot$. Our choices in age reflect significant phases in the evolution of WRs and ULXs, while our choices in metallicity largely align with the X-ray luminosity functions available for XRBs.

**ULXs** - We follow the methodology outlined in Garofali et al. (2024), which details the evolution of X-ray binaries (XRBs) as a function of stellar population metallicity and age. We focus on modeling ULXs, a subclass of XRBs that are a favorable choice for providing an upper limit on the X-ray contribution to the aggregate light from dwarf galaxies, as outlined below.

XRBs contain a star and a compact object in a binary system, undergoing mass transfer and accretion from the donor star to the compact object. HMXBs are a subclass of XRB that contains a high mass donor star ($> 8 M_\odot$) and form on short timescales ($< 5$ Myr) after a starburst (Linden et al. 2010; Wiktorowicz et al. 2017; West et al. 2023). This delay allows time for the primary star to evolve and form a compact object, as well as the secondary to evolve sufficiently for the onset of mass transfer. As a result, HMXBs emerge as the dominant sources of galaxy-integrated X-ray emission in scenarios without a central accreting supermassive black hole, especially in stellar populations younger than 100 Myr (Fragos et al. 2013; Lehmer et al. 2016). This makes

HMXBs relevant for dwarf galaxies, which often have bursty SFHs (Lee et al. 2007; Ting & Ji 2024).

Relatedly, a ULX is an XRB that emits intense X-ray radiation at the bright end of the HMXB luminosity function ($L_X \gtrsim 1 \times 10^{39}$ erg s$^{-1}$), which is a consequence of super-Eddington mass ($\sim$5 Eddington) supply rates onto a stellar mass compact object.

Due to energy conservation, super-Eddington accretion indicates the formation of powerful outflows or winds (King et al. 2001; Gladstone & Roberts 2009; King et al. 2023). Disk winds contributing to the EUV SED differentiate a ULX from an HMXB (Garofali et al. 2024). ULX models include an additional disk wind component for the accretor (Pinto et al. 2016; Kosec et al. 2018), more complex than multi-color blackbody models often used for HMXBs (e.g., Mitsuda et al. 1984; Trani et al. 2014; Senchyna et al. 2020). Consequently, ULXs can provide an upper bound on the hardness of the XRB ionizing continuum. Recent work suggests possible differences in ULX SED in the EUV regime tied to intrinsic source properties (Gúrpide et al. 2024). In this work, we are interested in accounting for the potential contribution from ULXs in the search for IMBHs and therefore do not consider ULX models with weaker EUV continua.

The well-established inverse correlation between the number of HMXBs or ULXs and galaxy metallicity is an additional favorable characteristic when considering these sources as the X-ray contaminants in the aggregate light from dwarf galaxies, which are low metallicity given their low stellar masses (e.g., Figure 1). Mapelli et al. (2010) shows an inverse relationship between the number of ULXs and the metallicity of the galaxies examined, indicating that a lower metallicity increases the formation of massive BHs, which in turn could power ULXs. Linden et al. (2010) reveals that low-metallicity environments allow for the formation of ULXs that undergo stable Roche lobe overflow with mild super-Eddington accretion, thereby increasing the quantity of these luminous sources compared to solar metallicity environments. The ULX SED *shape itself* does not vary with age, but age affects its normalization due to the changing ULX population over time. Metallicity influences both the shape and normalization of the ULX SED, as it affects both the formation rate of ULXs and their spectral properties.

We therefore use ULXs with massive donor stars to establish an upper limit for the non-IMBH contribution to X-ray excitation in dwarf galaxies. Because the neutron star population in XRBs is low on short time scales, we use BHs for the compact object component. We address the possibility of IMBHs serving as the compact object component of ULXs in Section 9.3. Our BHs are modeled at the following masses at each metallicity: 8 $M_\odot$ for $Z \geq 0.1$ $Z_\odot$ and 16 $M_\odot$ for $Z < 0.1$ $Z_\odot$ (Wiktorowicz et al. 2017; Garofali et al. 2024), however, this does not strongly affect the shape of the resulting SED. Our BH masses vary only as a function of metallicity, for there is no concrete evidence for the $M_\odot$ varying as a function of age. We use the post-starburst ages of the stellar models: 1, 2, 3.16, 5.01, 7.94, 10, 15.8, 20, 25.1 Myr and metallicities $\log Z/Z_\odot = -2.0, -1.3, -1.0, -0.7, -0.4, 0.0$. The solid lines in Figure 3 show a steady increase in X-ray luminosity with decreasing metallicity and a very slight shift in the location of the accretion disk's "big blue bump" due to the subtle difference in black hole mass.

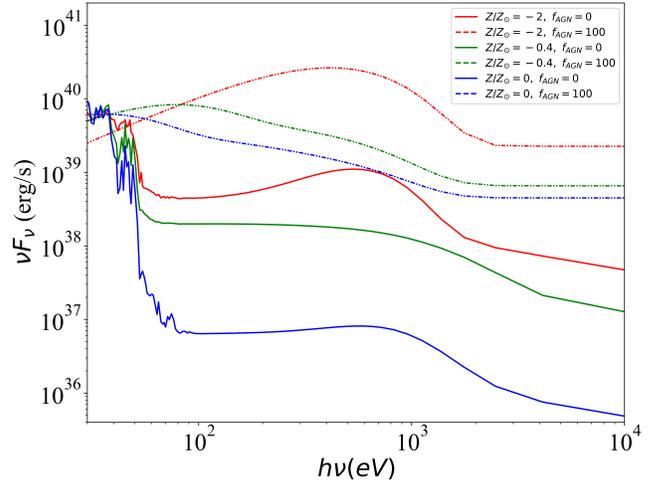

**Figure 3.** Spectral energy distributions with $\sim$2000 data points for $\log Z/Z_\odot = 0.0, -0.4, -2.0$ assuming a light BH seeding channel so that $\log M_{BH}/M_\odot = 5.89, 4.59, 2.0$ for each metallicity, respectively. The model with $f_{AGN} = 0\%$ has a post-starburst age of 25.1 Myr (maximum ULX contribution relative to stars) and the model with $f_{AGN} = 100\%$ has a post-starburst age of 1 Myr (no ULX contribution).

**IMBHs** - We follow the methodology from Richardson et al. (2022), which focuses on advancing the understanding of IMBHs in the mass range of $10^3 - 10^5 M_\odot$ by addressing significant uncertainties in photoionization modeling. Studying three different AGN SED models, the selected models were the `qsosed` (Panda et al. 2019; Sarkar et al. 2021) and "disk-plaw" (Mitsuda et al. 1984).

The disk-plaw SED (Cann et al. 2019; Bhat et al. 2020) combines the multi-color blackbody accretion disk model (Mitsuda et al. 1984) with a power law normalized to give a specific $\alpha_{ox}$, but lacks a physical connection between the parameters. Although disk-plaw is useful for AGN photoionization modeling due to its physical motivation, it relies upon empirical relations calibrated from observations of giant AGN that may or may not apply to dwarf AGN. The `qsosed` model is self-consistent and accounts for the inner accretion disk radius required to power the X-ray emission. This model contains a small number of free parameters and directly calculates values like $\alpha_{ox}$ based on fundamental BH properties such as mass, spin, and inclination.

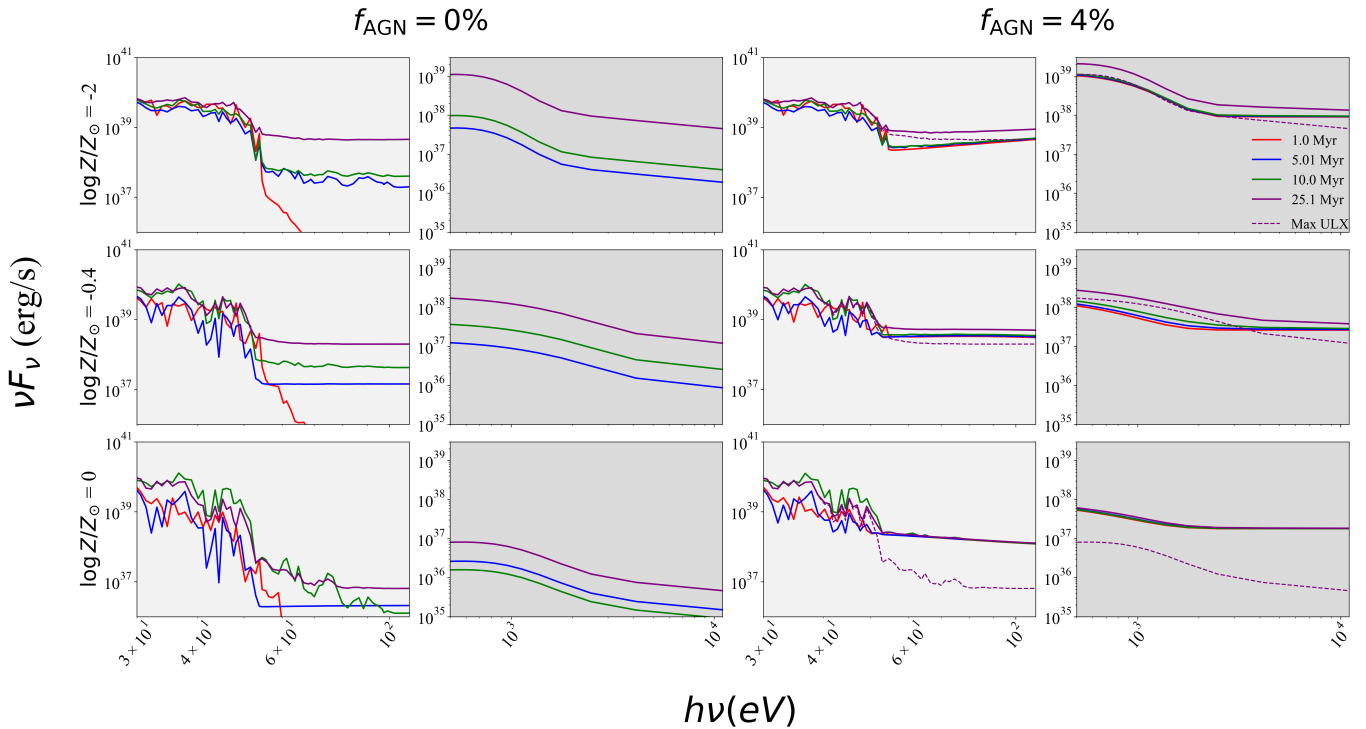

**Figure 4.** Spectral energy distributions for four distinct post-starburst ages (5.01 Myr - 25.1 Myr) at various metallicities ($\log Z/Z_\odot = 0.0, -0.4, -2.0$) and $f_{\rm AGN} = 0\%$ and $4\%$ assuming a light BH seeding channel. The purple dashed line indicates the maximum ULX contribution (post-starburst age = 25.1 Myr and $f_{\rm AGN} = 0\%$).

The IMBH SED is not well-constrained observationally. Therefore, we opt for a theoretical model anchored in the physics of accretion onto a BH, and we further opt for this model to have as few free parameters as possible. To this end, we select the `qsosed` for our model, which is a simplified version of the `agnsed` model from Kubota & Done (2018) with many parameters fixed to their "typical" values as determined by analysis of SMBHs.

The accretion rate in IMBHs is highly stochastic and shows several orders-of-magnitude variations throughout cosmic time (Bellovary et al. 2019). Since super-Eddington accretion is unlikely at $z < 1$, we assume an accretion rate of $\dot{m} \equiv \dot{m}/\dot{m}_{\rm edd} = 0.1$ with the `qsosed`. We do not incorporate models with highly sub-Eddington accretion rates since we seek to assess the emission line predictions generated by observable BHs. Nevertheless, we explore a small subset of models with $\dot{m} = 0.022$ and $\dot{m} = 0.5$ in Section 9.3. For simplicity, we assume that the dimensionless spin parameter $a_* = 0$, noting that black hole spin has a relatively small effect on the overall IMBH SED until reaching much higher values around $a_* \sim 0.8$.

In summary, the parameters for `qsosed` model are as follows: $\dot{m} = 0.1$, $a_* = 0$, and $i = 45°$. We include black hole masses ranging from $10^2 \, M_\odot$ to $10^{5.89} \, M_\odot$ to satisfy the $M_{\rm BH}$ vs. $Z$ relation in Section 2.3 for all three seeding mechanisms: heavy, moderate, and light. The dashed lines in Figure 3 show the variation in the IMBH SED as a function of metallicity, and therefore black hole mass, assuming a light BH seeding channel. The X-ray luminosity for IMBH SEDs spans a much smaller range than the ULX SEDs, but shows a much more prominent shift in the location of the accretion disk peak due to black hole mass spanning $10^2$ - $10^{5.89}$ $M_\odot$.

### 3.1.2. *Combined SEDs*

Combining the SEDs for excitation sources presented above results in an SSP+ULX model, which combines the stellar and ULX models at corresponding post-starburst ages and metallicities. This addition creates an SSP+ULX model that combines the SED characteristics of stellar populations and ULXs, ensuring that the physical properties, such as the timing of XRBs emergence and metallicity effects on stellar evolution and wind properties, are consistently represented across all ages. Fragos et al. (2013) includes simulations for only three metallicities (0.1, 1.0, and 1.5 $Z_\odot$). To create a more comprehensive model, we have interpolated between these values, which assumes that the age evolution between interpolated points is similar to that of nearby metallicities for which simulations were run. This means some specific age-related variations between metallicities may be missed. However, this approach is suitable for capturing overall trends with age and metallicity, which aligns with our primary goal: assessing ULX contributions to detect IMBHs across different post-starburst ages, metallicities, and AGN fractions. Providing more detailed ULX evolution models is beyond the scope of our study, especially given the lack of newer empirical or theoretical work on ULX age and metallicity evolution.

To create a composite SED, we add the IMBH SED to the SSP+ULX SED using an ionizing continuum with different $f_{\rm AGN}$, representing the fraction of the total ionizing photons attributed to the AGN SED, thereby constructing the overall SED shape. We vary the black hole mass of the `qsosed` model according to the $M_{\rm BH}$ vs. $Z$ relation presented in Section 2.3 for three different seeding channels: heavy (direct collapse), moderate (stellar mergers), and light (Population III stars).

The composite SSP+ULX+IMBH include models of nine different post-starburst ages (1, 2, 3.16, 3.98, 5.01, 7.94, 10, 15.8, 20, 25.1 Myr) with varying metallicities (log $Z/Z_\odot$ = $-2.0, -1.3, -1.0, -1.3, -0.7, -0.4, -0.15, 0.0$) and AGN fractions ($f_{\rm AGN}$ = 0.0, 0.02, 0.04, 0.08, 0.16, 0.32, 0.5, 0.64, and 1.0). Figure 4 displays a subset of SEDs from these models. Each double column represents a different $f_{\rm AGN}$ of 0.0 and 0.04, and each subcolumn represents a different range of photon energies. The light gray column highlights the UV, while the dark gray column shows the X-ray regime. Each row represents different metallicities log $Z/Z_\odot$= -2.0, -0.4, 0.0 in increasing order from top to bottom. Due to a Bayesian inference conducted in Polimera et al. (2022), we selected $Z/Z_\odot = 0.4$ as a fiducial metallicity for $z \sim 0$ dwarf galaxies. The other metallicities correspond to the minimum and maximum values considered in our models.

In Figure 4, each color signifies a different post-starburst age. The "max ULX" line at each metallicity in the $f_{\rm AGN} = 0.04$ columns shows the ULX contribution at 25.1 Myr (timescale from our grids at which the ULX has the maximum relative contribution to the stellar ionizing continuum) and $f_{\rm AGN} = 0$, indicating the maximum X-ray contribution from ULX without AGN contributions at each metallicity.

At 1 Myr and $f_{\rm AGN} = 0.0$, ULXs are not contributing to the composite spectrum by construction. In Figure 4, the 1 Myr model (displayed by the red line) is absent in the X-ray, $f_{\rm AGN} = 0.0$ subcolumn. ULX contributions start at 5 Myr, displaying the ULX dependence on age. At 5 Myr, O/WR stars have enough time to die off and a stellar mass BH can begin accreting. However, due to the AGN contribution, the 1 Myr model is present in the $f_{\rm AGN} = 0.04$ subcolumn. The IMBH SED is not dependent on age, and its contribution is only affected by the $f_{\rm AGN}$ and $Z$ parameters set for each model. In Figure 4, the other three age models ($\geq 5$ Myr) are present in the X-ray subcolumn of $f_{\rm AGN} = 0.0$. As massive stars deplete over time, the stellar ionizing continuum weakens, but ULX production continues as compact objects form and accrete on evolutionary timescales tied to the companions. This creates the appearance of a more substantial relative contribution from ULXs to the ionizing continuum at later stages.

Metallicity changes the SED by affecting both the ULX and the IMBH. Higher metallicities will have a lower ULX population and a slightly lower ULX black hole mass, translating to a lower luminosity but harder X-ray continuum. In contrast, higher metallicity is correlated with higher IMBH mass (Section 2.3), which softens the continuum, but the IMBH contribution to $L_X$ mainly depends on $f_{\rm AGN}$.

The overall effect can be seen by comparing the models between solar metallicity and log $Z/Z_\odot = -2$ in Figure 4 where all of the solar metallicity SEDs have less flux in X-ray regime. The same effect is present in the EUV for $f_{\rm AGN}$= 0, but the IMBH component for $f_{\rm AGN}$= 0.04 holds the EUV flux relatively constant across all metallicities. The bottom row of Figure 4, which is at solar metallicity, indicates that an AGN contribution of 4% is enough to completely erase any ULX signature at energies >54 eV.

Figure 5 shows that at low metallicities, ULXs continue to contribute to the ionizing spectrum at >54 eV until the AGN contribution reaches 16%, but the dependence on post-starburst age weakens significantly. This is because the IMBH dominates the X-ray spectrum at higher $f_{\rm AGN}$ and high metallicities. Figure 5 emphasizes that the metallicity affects the shape of the EUV spectrum at energies $\geq 50$ eV as evident when comparing the left column at $Z/Z_\odot = -2$ and to the right column at $Z/Z_\odot = -0.4$. However, the spectrum at energies $\leq 30$ eV continues to show substantial variation with post-starburst age in the $Z/Z_\odot = -0.4$ case, which is where ions like Ar II, Ne II, and S III are formed.

The discrepancies between different SEDs at low metallicities in Figures 4 and 5 at any $f_{\rm AGN}$ motivate a more comprehensive dwarf galaxy model when identifying IMBHs. These results have implications for using emission line diagnostics sensitive to high energy photons, like Ne V, He II, and O I, for separating IMBHs and ULXs contributions at various metallicities in dwarf galaxies.

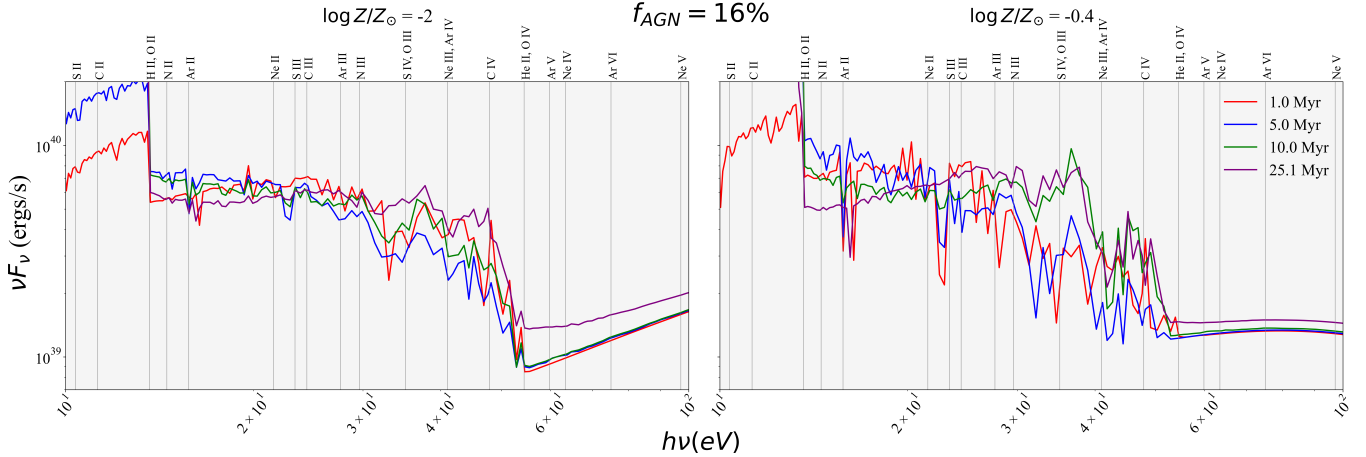

**Figure 5.** Each panel shows multiple SEDs with different post-starburst ages of 1.0, 5.01, 10.0, and 25.1 Myr. The left panel has a metallicity set at $\log Z/Z_\odot = -2$. The right panel has a metallicity of $\log Z/Z_\odot = -0.4$. The AGN contributes 16% to the total energy output, as denoted by $f_{\rm AGN} = 16\%$, and assumes a light BH seeding channel. Both panels have vertical gray lines that represent relevant ionization potentials of different elements, as indicated by the labels above the graph, such as S II, C II, O II, etc.

## 3.2. Cloud Physical Conditions

Indirectly detecting IMBHs through ISM emission line signatures not only depends upon robust SED modeling but also attention to detail with regard to cloud-based physics. The emission line predictions reflect observations only as well as we can model the excitation source and processes occurring in the ionized cloud. For the cloud, we apply the same physical conditions as described in Richardson et al. (2022), which we briefly summarize in this section, using the photoionization code Cloudy v17 (Ferland et al. 2017). We select a hydrogen density of $\log n_{\rm H} = 2.0$ at the illuminated face and ionization parameters ranging $-4.0 \leq \log U \leq -1.0$ in increments of 0.25 dex. We include a magnetic field of $B_0 = 100$ $\mu$G with $\kappa = 2/3$ and a small amount of turbulence ($v = 2$ kms$^{-1}$) as in Abel et al. (2009). We assume constant total pressure holds until the simulations stop at $n_e/n_{\rm H} = 0.01$. We use the nebular abundances and scaling from Nicholls et al. (2017) where the solar values are known as Galactic Concordance abundances (12+log (O/H) = 8.76), largely based on local B-star abundances. We assume that Orion grains (Baldwin et al. 1991) and polycyclic aromatic hydrocarbons (PAHs, Abel et al. 2008) are present and scaled according to the dust-to-gas ratios presented in Rémy-Ruyer et al. (2014). Gas-phase abundances are depleted using a custom set of depletion factors largely based on Jenkins (2009) and Jenkins & Wallerstein (2017), specifically provided in Richardson et al. (2022), and similar to the depletion factors derived in Gunasekera et al. (2022). We assume a closed geometry with co-incident mixing of the ionizing continua (Richardson et al. 2022). Running simulations with other geometries and mixing methodologies are beyond the scope of this work, however, we discuss the implications of using an open geometry and non-coincident mixing throughout the paper based on the results from Richardson et al. (2022).

## 4. LINE RATIO SENSITIVITY TO BLACK HOLE SEEDING AND $M_{\rm BH}$

Figures 6-8 feature many common emission line ratios used to identify AGN excitation in the UV through IR. The panels in each figure show the sensitivity of these line ratios to $\log Z/Z_\odot$ (bottom axis), and therefore $\log M_{\rm BH}/M_\odot$ (top axis), assuming a light seeding channel and a post-starburst age of 1 Myr (e.g., no ULX contribution). These emission line ratios are normalized to the predictions from a model with a fiducial IMBH mass of $10^5$ $M_\odot$. The normalization in the figures allows the comparison between heavy seeding and other seeding channels since the slope of any line segment in a given panel relates to sensitivity to seeding, with slopes closer to zero signaling insensitivity to seeding. At $Z/Z_\odot \lesssim 0.6$ the heavy seeding model flattens to a constant value of $M_{\rm BH} = 10^5$ $M_\odot$ (Figure 2), and any difference in the emission line predictions at lower metallicities would be due to moderate or light BH seed channels. Similarly, emission line ratios that continue to evolve at $Z/Z_\odot \lesssim 0.1$, where the runaway stellar merger chan-

nel flattens to a constant value of $M_{BH} = 10^{3.5}$ M$_\odot$, are sensitive to the light seeding channel (dark gray boxes). The vertical stratification of the different linestyles and colors shows sensitivity to $\log U$ and $f_{AGN}$, respectively (discussed in more detail in Section 3).

We note that our assumption of a 1 Myr post-starburst age is meant to isolate the effects of changing $M_{BH}$ and seeding mechanism, and exclude ULX contributions, which could be used to identify emission line ratios that could constrain these properties. In reality, ULXs are present in galaxies at least to some degree, which means the results in this section are not meant to represent the emission from actual galaxies.

### 4.1. UV

Figure 6 provides many emission line ratios used as diagnostics for separating AGN and SF activity in the UV (Feltre et al. 2016; Jaskot & Ravindranath 2016; Nakajima et al. 2018; Cleri et al. 2023; Garofali et al. 2024; Mingozzi et al. 2024). All ratios show greater sensitivity (y-axis for a given $Z$) to $M_{BH}$ and seeding channel as the ionization parameter increases, however the overall sensitivity across the full $M_{BH}$ and $Z$ range is rarely greater than 0.3 dex. This suggests that photoionization models assuming a constant $M_{BH}$ that predict these emission line ratios should account for a factor of two uncertainty due to variable IMBH mass.

The one exception is [Ne V] $\lambda 3437$/[Ne III] $\lambda 3869$ (Cleri et al. 2023), which drastically changes with $M_{BH}$ and $Z$. The models at $\log U = -3.5$ yield a [Ne V] luminosity that is likely undetectable (Section 5.1), therefore the most relevant trends at $\log U \geq -2.5$ show $\approx 1.5$ dex deviation in [Ne V]/[Ne III]. Consequently, photoionization models must account for seeding channels and changing black hole mass to make proper comparisons to dwarf AGN with [Ne V]/[Ne III] observations (Cleri et al. 2023; Chisholm et al. 2024). Similarly, [Ne V]/[Ne III] provides strong diagnostic value in inferring the $M_{BH}$ and seeding channel in dwarf galaxies since [Ne III] and [Ne V] have widely separated ionization potentials and modest excitation energies.

### 4.2. Optical

Figure 7 provides many emission line ratios used as diagnostics for separating AGN and SF activity in the optical (Kewley et al. 2001; Lamareille 2010; Shirazi & Brinchmann 2012; Backhaus et al. 2022; Mazzolari et al. 2024; Mingozzi et al. 2024). In general, as both $U$ and $f_{AGN}$ increase, the black hole mass shows greater influence on the emission line predictions. Most of the emission line ratios present modest variations with $M_{BH}$ only growing to $\sim 0.2$ dex at the lowest metallicity. The notable exceptions to this trend are ratios including He II $\lambda 4686$, [O I] $\lambda 6300$, and [S II] $\lambda 6720$, which vary $\sim 0.5 - 0.8$ dex across the full range of metallicities.

All three of these line ratios have been shown as effective indicators for identifying giant AGN (Veilleux & Osterbrock 1987; Kewley et al. 2006; Shirazi & Brinchmann 2012; Feltre et al. 2023) and dwarf AGN (Sartori et al. 2015; Polimera et al. 2022; Mezcua & Domínguez Sánchez 2024). Our results emphasize the need to vary $M_{BH}$ when comparing models to observations with these line ratios. Together, they could provide valuable constraints on black hole mass and seeding channels for different reasons. The high ionization potential of He II (54 eV) effectively traces the thermal peak of the accretion disk in the EUV, while the [S II] $\lambda 6720$ and [O I] $\lambda 6300$, which form close to the ionization front, effectively trace the soft X-rays that freely travel through H II region until encountering neutral gas. The emission lines from He II and [O I] additionally trace $f_{AGN}$ quite well at very low metallicities (Section 5.2) while remaining resistant to changes due to gas cloud geometry and SED mixing methodology (Richardson et al. 2022).

### 4.3. IR

Figure 8 shows that emission line ratios in the mid-IR mostly show the same trend as the optical: greater $U$ and $f_{AGN}$ leads to greater differences in emission due to black hole mass. In contrast, however, the variation on account of $M_{BH}$ is far larger than the UV and optical despite the ionization potentials of the emission lines being reasonably similar across all wavelength bands (see Section 4.4 for discussion). All the emission line ratios shown could effectively trace black $M_{BH}$ and seeding channel, although degeneracies exist. For example, the predictions for [S IV] $10.5\mu$m /[Ar II] $6.99\mu$m using a fiducial black hole mass or variable black hole mass are roughly equivalent at $Z = 0.01$ Z$_\odot$ and $Z =$ Z$_\odot$ for $\log U = -3.5$. In most situations, global galaxy properties such as stellar masses could assist in distinguishing between the two cases, but it does serve as a warning about making strong conclusions with few emission line tracers or galaxy properties. Notably, [Ne V] $15.56\mu$m/[Ne III] $14.2\mu$m and [Ar II] $6.99\mu$m/Pf$\alpha$ are largely not degenerate. However, as with [Ne V] in the optical, the [Ne V] $14.2\mu$m emission line only becomes detectable for high $f_{AGN}$ or high $U$, uncharacteristic of the general dwarf galaxies population (Section 5.1). For $U > -2.5$, line ratios featuring [O IV] $25.9\mu$m and [Ne II] $12.8\mu$m trace $f_{AGN}$ well without any degeneracy. Additionally, the emission line ratios [O IV]/[Ne III] and [Ar II]/Pf$\alpha$ are largely insensitive to gas cloud geometry and SED mixing method-ology (Richardson et al. 2022), which coupled with the results in Figure 8, makes them prime candidates to unveil black hole masses and seeding mechanisms.

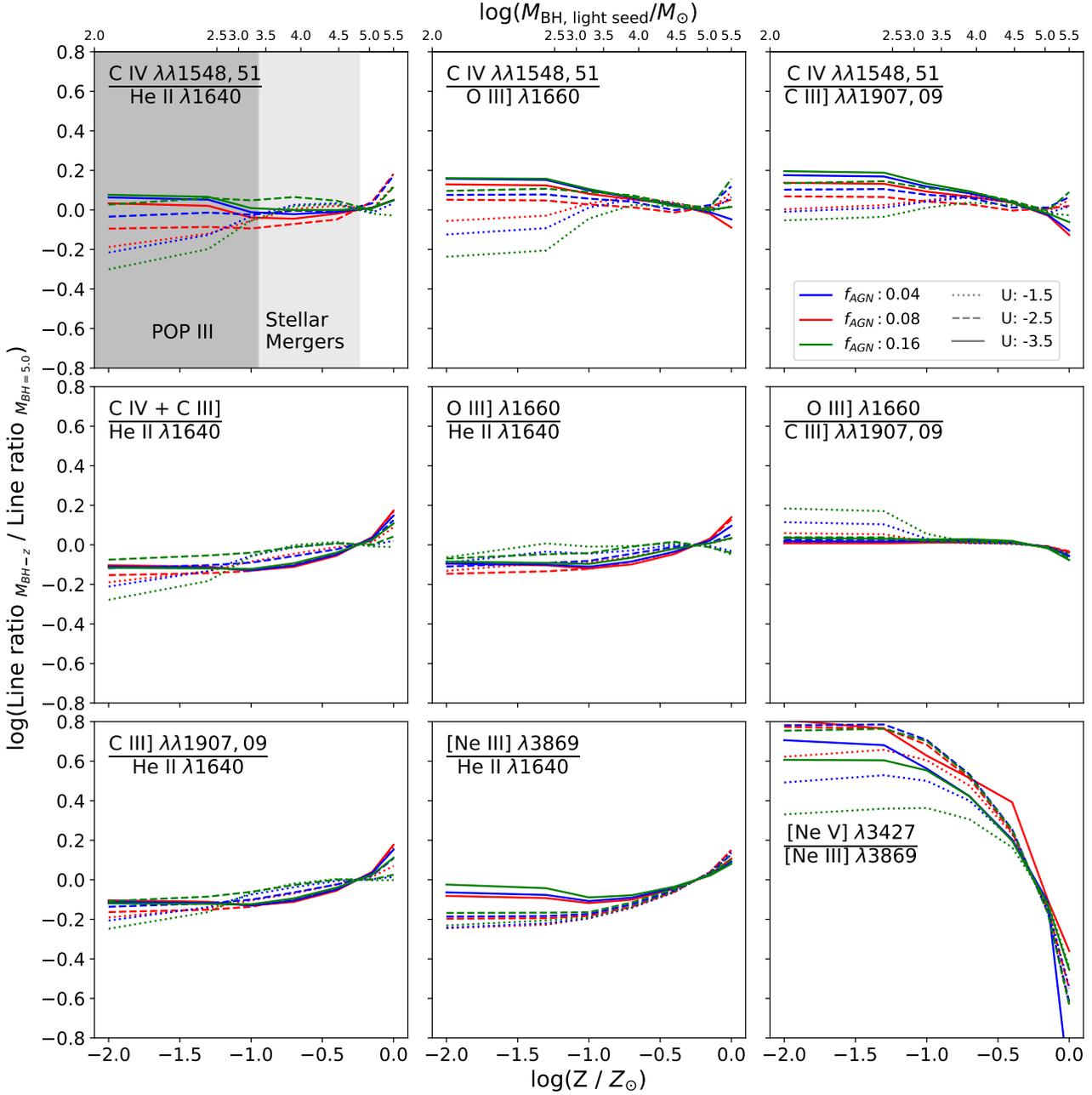

**Figure 6.** Emission line ratio sensitivity to $Z$ (bottom axis) and $M_{\rm BH}$ (top axis) for several common UV AGN diagnostics assuming a 1 Myr post-starburst age (no ULX contribution). The emission line predictions are provided for the light seeding channel relative to a fiducial model assuming $M_{\rm BH} = 10^5$ $M_\odot$. The heavy, moderate, and light seeds reach an $M_{\rm BH}$ floor at $\log M_{\rm BH}/M_\odot = 5.0, 3.5$ and $2.0$, respectively. The gray boxes represent regions where different seed mechanisms start to take effect. Except for [Ne V]/[Ne III], most of the line ratios in the UV are insensitive to $M_{\rm BH}$ and IMBH seeding channels.

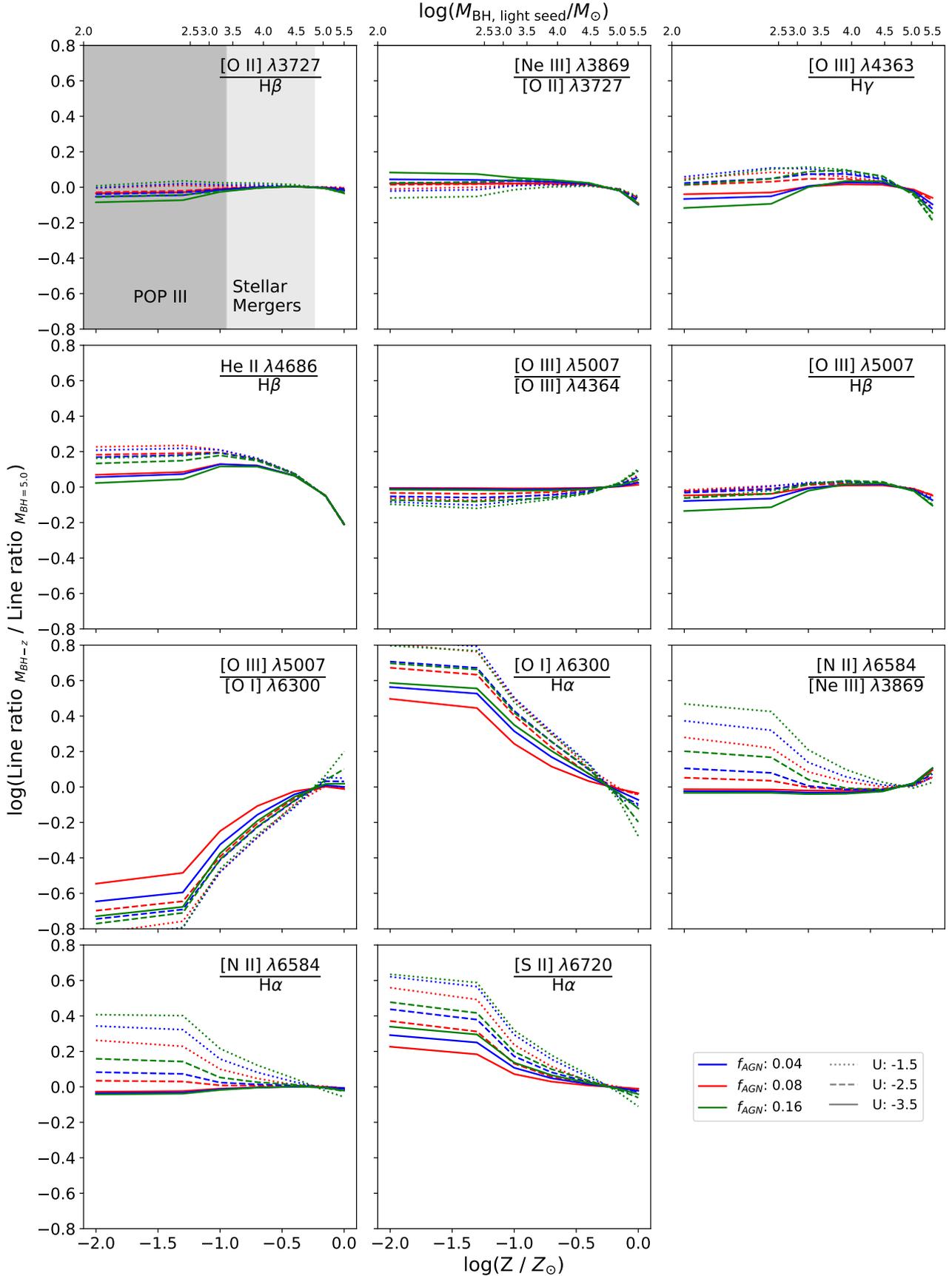

**Figure 7.** Sensitivity to $Z$ and $M_{BH}$ for several common optical emission line ratio AGN diagnostics in the same manner as those from Figure 6. Emission line ratios that contain He II $\lambda 4686$, [O I] $\lambda 6300$, and [S II] $\lambda 6720$ show the greatest sensitivity, which makes them suitable candidates for constraining $M_{BH}$ and seeding channels.

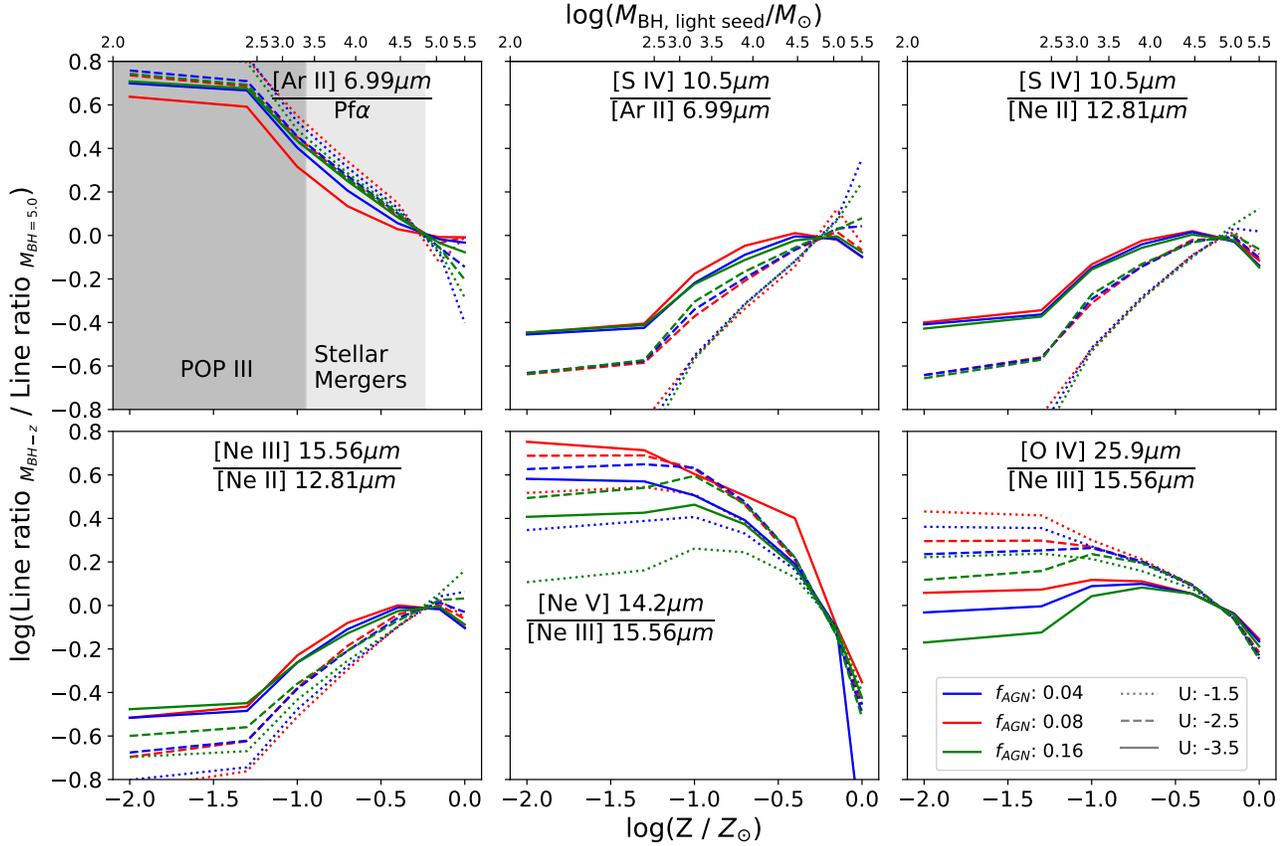

**Figure 8.** Sensitivity to $Z$ and $M_{BH}$ for several common optical emission line ratio AGN diagnostics in the same manner as those from Figure 6 assuming a 1 Myr post-starburst age (no ULX contribution). IR emission line ratios are more sensitive than both UV and optical emission line ratios with [S IV] $\lambda 10.5\mu$m, [Ne II] $\lambda 12.8\mu$m, and [Ar II] $\lambda 6.99\mu$m showing the greatest sensitivity. Each subplot that contains one or more of these wavelengths shows a maximum difference of $\gtrsim 0.8$ dex.

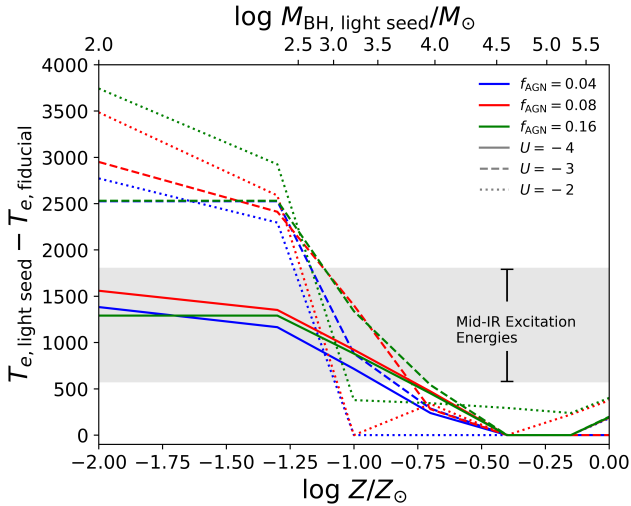

**Figure 9.** The difference between $T_e$ at the face of the cloud for the light seeding model and the fiducial model with $M_{BH} = 10^5 M_\odot$. The variations in $M_{BH}$ associated with the light seeding model introduce differences in $T_e$ on the order of the emission line excitation energies in the mid-IR.

### 4.4. *Interpreting $M_{BH}$ Sensitive Line Ratios*

Decreasing $M_{BH}$ hardens the IMBH SED since the accretion disk reaches higher temperatures. Therefore, to first order, line ratios sensitive to $M_{BH}$ and seeding should either include ions with widely separated ionization potentials (e.g., [Ne V]/[Ne III]) or low ionization / neutral ions (e.g., [O I]/H$\alpha$, [Ar II]/Pf$\alpha$), which are sensitive to X-ray photons penetrating neutral gas. Indeed, Figures 7 and 8 show that [Ne V]/[Ne III], [O I]/H$\alpha$, and [Ar II]/Pf$\alpha$ are very sensitive to changes in $M_{BH}$ and the assumed seeding mechanism. However, Figures 6 and 8 display line ratios with similarly large differences in ionization potential (e.g., O III]/He II compared to [O IV]/[Ne III]), and yet the UV lines are much less sensitive to $M_{BH}$ than the IR lines. This emphasizes a secondary effect: the electron transitions in the UV require higher $T_e$ than the electron transitions in the IR. Most of the lines in Figures 6-8 are collisionally ex-cited by electrons. Figure 9 displays the *difference* in $T_e$ that occurs from assuming a light seeding channel instead of fiducial $M_{BH}$ as a function of $Z$ and $M_{BH}$. The variation in $T_e$ between seeding channels is on the order of the excitation energies in the mid-IR, explaining the sensitivity of the lines to $M_{BH}$ and seeding. In contrast, the excitation energies in the optical start at ∼8000 K, which explains the weaker sensitivity to $M_{BH}$ and seeding in the optical and UV.

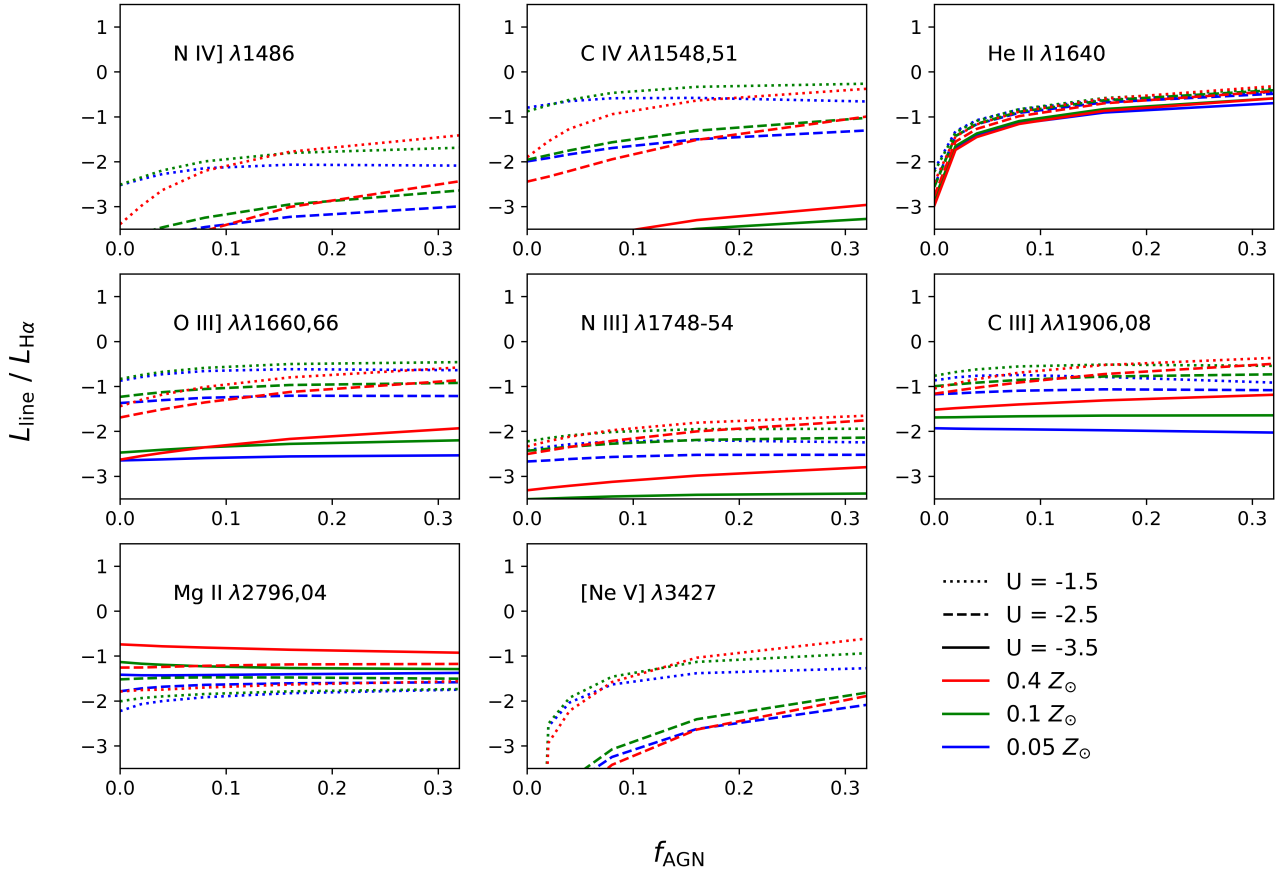

**Figure 10.** The luminosities of potential AGN diagnostics in the UV relative to Hα assuming a post-starburst age of 1 Myr (i.e, no ULX contribution).

## 5. DETECTING DIFFERENCES IN AGN FRACTION

Emission line ratios resulting from two ions with a large difference in ionization potential or including low ionization lines tend to be sensitive to changes in $f_{AGN}$. However, the detectability of emission lines depends upon the physical conditions present and instrument sensitivity, meaning good $f_{AGN}$ tracers might not be observable. We address this topic in Figures 10-12 by scaling all emission line luminosities to the Hα luminosity, which is essentially independent of $f_{AGN}$ for a given value of ionization parameter. By choosing a reference standard across all wavelengths, we avoid selecting arbitrary distances needed to compare our predictions to the line luminosity sensitivities of various instruments.

Indeed, Hα should be detectable across the full range of black hole masses incorporated into our models at non-zero redshifts. As an example, the dwarf AGN in the MANGA survey from Mezcua & Domínguez Sánchez (2024) show bolometric luminosities as low as $L_{bol} = 10^{38}$ erg/s at $z < 0.15$. Assuming an accretion rate of 10% the Eddington limit, this translates to a 10 solar mass BH. Given the strong Hα measured across the Mezcua & Domínguez Sánchez (2024) sample, the $M_{BH} = 10^2 - 10^{5.8} M_\odot$ range assumed in our simulations should produce Hα emission within the limits of detection.

In Figures 10-12, the line color represents metallicity while the linestyle represents the ionization parameter. Flatter lines are insensitive to $f_{AGN}$, and stratification of line styles denotes sensitivity to log $U$ (i.e., some lines only work for high ionization environments.)

### 5.1. *UV*

Figure 10 shows the luminosities of UV emission lines relative to Hα used for identifying AGN as a function of $f_{\rm AGN}$. While N IV] λ1486 and [Ne V] λ3427 show ∼1.0 dex variations with $f_{\rm AGN}$, the lines are likely only detectable at high ionization log $U \approx -1.5$ for the lowest $f_{\rm AGN}$, and at intermediate ionization $U \approx -2.5$ with a higher $f_{\rm AGN}$. Variations in C IV λλ1548,51 with $f_{\rm AGN}$ range from 0.1 dex to 1.0 dex depending on both ionization and metallicity, but C IV λλ1548,51 formed in low ionization gas is likely difficult to detect. He II λ1640 shows ∼ 0.5 dex variation with $f_{\rm AGN}$ but remains detectable over a wide range of ionization and metallicity. The other emission lines in Figure 10 are detectable with intermediate to high ionization across the entire $f_{\rm AGN}$ range; however, they do not effectively trace $f_{\rm AGN}$.

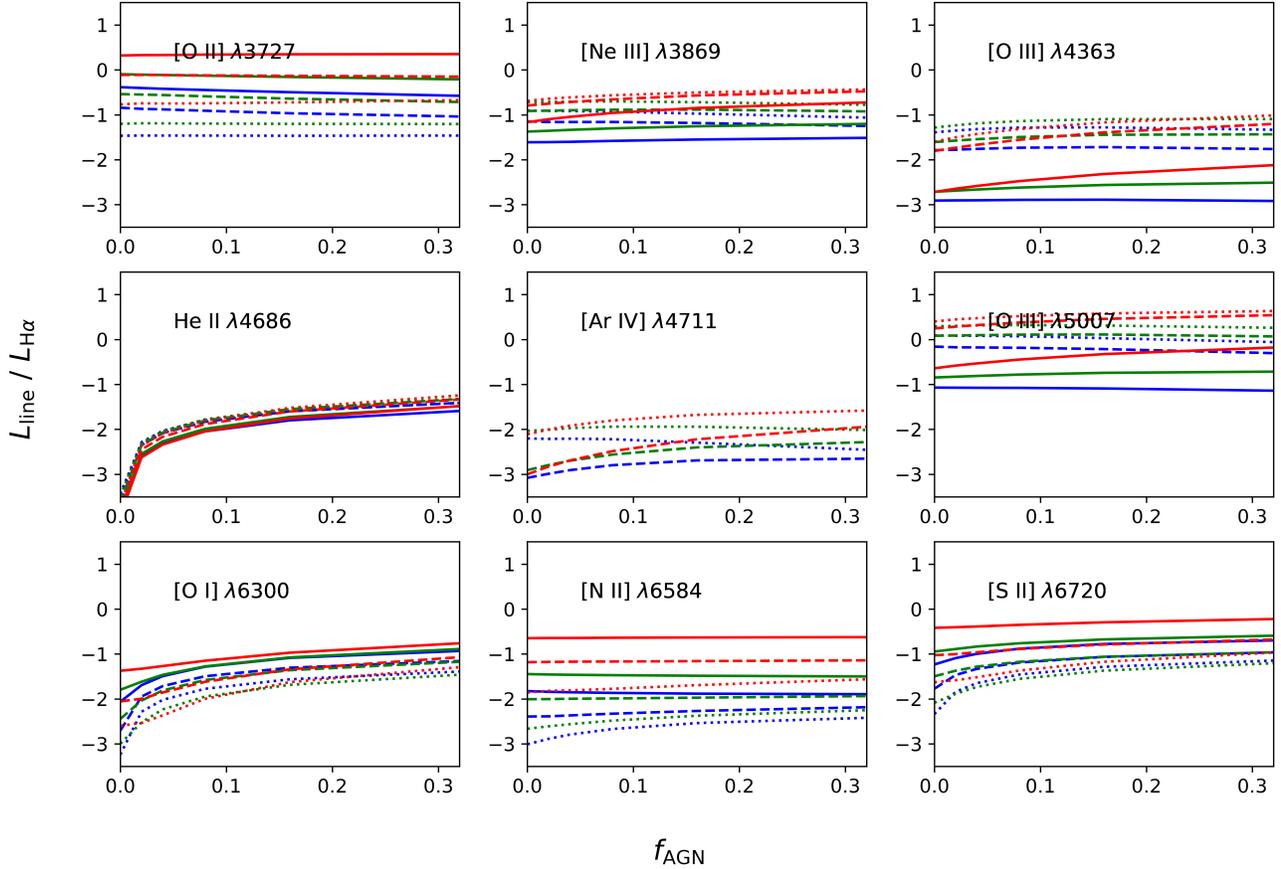

**Figure 11.** The line luminosities of potential AGN diagnostics in the optical relative to Hα assuming a post-starburst age of 1 Myr (i.e, no ULX contribution). Linestyles and colors are the same as Figure 10.

### 5.2. *Optical*

Figure 11 shows that many optical emission lines are insensitive to $f_{\rm AGN}$ despite being easily detected across a range of physical conditions. Notably, the lines that construct the BPT diagram and the "OHNO diagram" ([O II] λ3727 and [Ne III] λ3869 (Backhaus et al. 2022) remain relatively flat with respect to $f_{\rm AGN}$. In contrast, He II λ4686, [O I] λ6300, and [S II] λ6720 all vary appreciably with $f_{\rm AGN}$. In particular, He II λ4686 shows a smooth arc of increasing luminosity up to 1.0 dex, independent of both ionization parameter and metallicity. The low ionization line [S II] λ6720 displays a greater range of luminosities than He II λ4686 but still varies by ∼1.0 dex with $f_{\rm AGN}$ for $Z/Z_\odot < 0.4$. The neutral line [O I] λ6300 traces $f_{\rm AGN}$ slightly better than [S II] λ6720 even at intermediate metallicities (Richardson et al. 2022).

### 5.3. *IR*

Figure 12 shows that luminosities of [Ar III] 8.99μm, [S IV] 10.5μm, [Ne III] 15.6μm, and [S III] 18.7μm do not appreciably vary with $f_{\rm AGN}$. As with [Ne V] λ3427 in Figure 10, [Ne V] 14.2μm traces $f_{\rm AGN}$ rather well at high ionization or at intermediate ionization if $f_{\rm AGN} > 0.16$. Emission lines with even higher ionization potentials than Ne V, like [Ar V] 7.90μm, are significantly weaker than [Ne V] 14.2μm and are likely more difficult to detect. In contrast, [O IV] 25.9μm has a slightly lower ionization potential than Ne V and traces $f_{\rm AGN}$ very well over a wide range of ionization (log $U \gtrsim -2.5$) and metallicity ($Z/Z_\odot > 0.01$). While [O IV] 25.9μm would be weak at low ionization, the emission lines [Ar II] 6.99μm and [Ne II] 12.8μm are stronger in this regime, and trace $f_{AGN}$ equally as well as [O IV] 25.9μm.

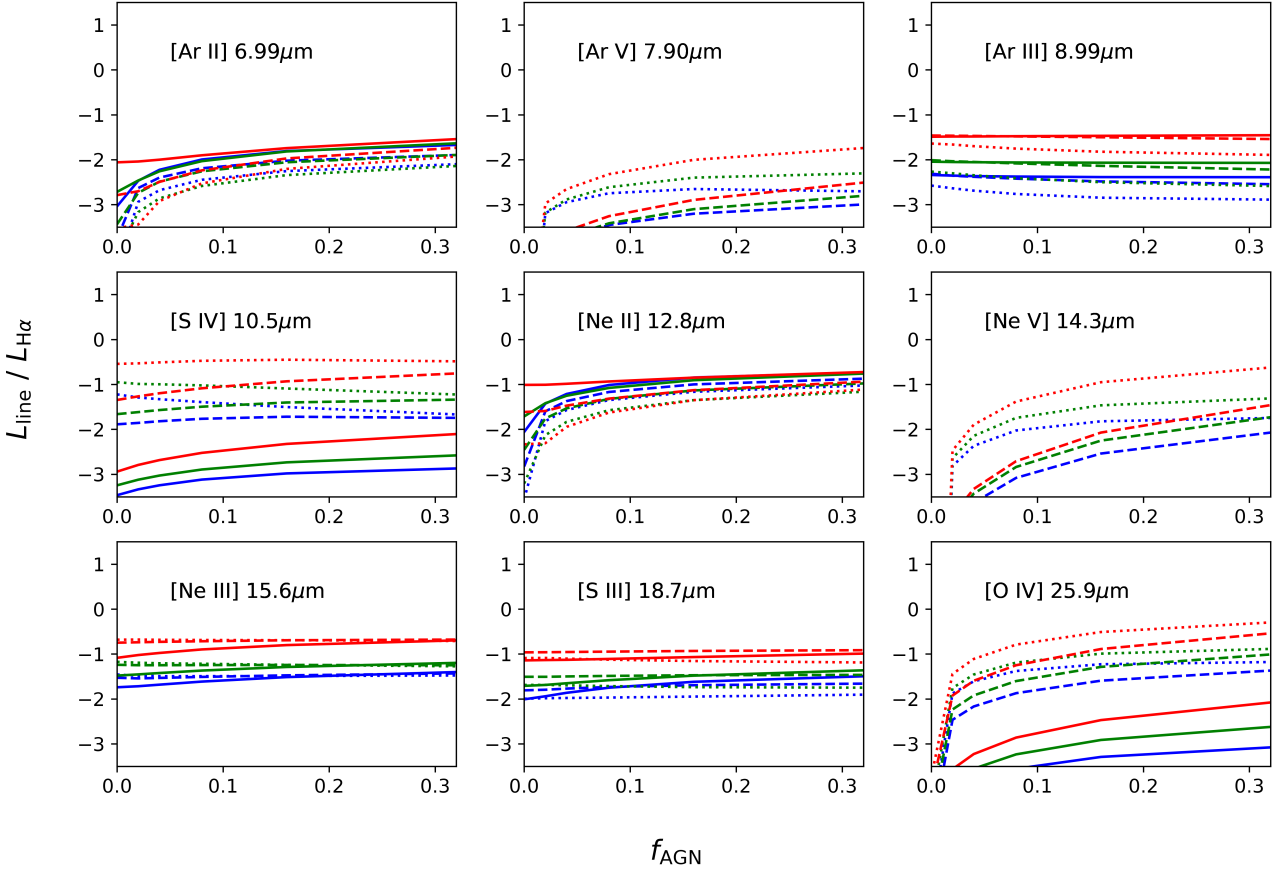

**Figure 12.** The line luminosities of potential AGN diagnostics in the IR relative to Hα assuming a post-starburst age of 1 Myr (i.e, no ULX contribution). Linestyles and colors are the same as Figure 10.

## 6. ASSESSING ULX CONTRIBUTIONS

### 6.1. IR-UV

ULXs can not only generate emission line signatures that mimic the presence of AGN (Berghea et al. 2010; Reyero Serantes et al. 2024) but also contribute significantly to the emission when an AGN is present. In this section, we quantify the influence of ULXs on AGN diagnostics when an active IMBH is also present. While the maximum X-ray luminosity due to a ULX occurs at 5 Myr when ULXs first form, the maximum ULX contribution to ionizing luminosity relative to stars occurs at 25.1 Myr (Garofali et al. 2024). Figure 13 displays the ratio of the predicted AGN diagnostics assuming 1 Myr and 25.1 Myr post-starburst ages and a light seed channel with the gold bar representing when the ULX contribution is lower than 10%. As $f_{AGN}$ increases, the ULX contribution to a given line ratio becomes negligible in most cases by $f_{AGN} \approx 0.32$, although deviations occur depending upon $U$, $Z$, and line ratio. In particular, He II λ4686/Hβ and line ratios including [O IV] 25.9 μm quickly approach contribution levels of ≈ 5% with little dependence on $U$ and $Z$, which makes them more resistant to the influence of a ULX compared to other AGN diagnostics. On the other hand, [Ne III] λ3869/[N II] λ6584 and [Ne III] 15.6μm/[Ne II] 12.8μm are particularly sensitive to the ULX at low metallicity, and line ratios including C IV λλ1548,50 show strong sensitivity to the ULX over a large range of $U$ and $Z$.

This result emphasizes the nuances resulting from the ULX SED varying with post-starburst age and $Z$, combined with the IMBH SED varying with $Z$, sometimes create large effects that should be systematically investigated. For physically realistic environments with a range of ISM conditions and potentially a mixture of sources (that depend on post-starburst age and metallicity), no single line ratio or diagnostic can robustly characterize the underlying conditions. Erroneous conclusions may result if one attempts to represent the excitation source using a single diagnostic and assumes the physical environment can be approximated by a simplistic model (e.g., single cloud, single source). We further explore the implications of this in Section 9.1.

18 RICHARDSON ET AL.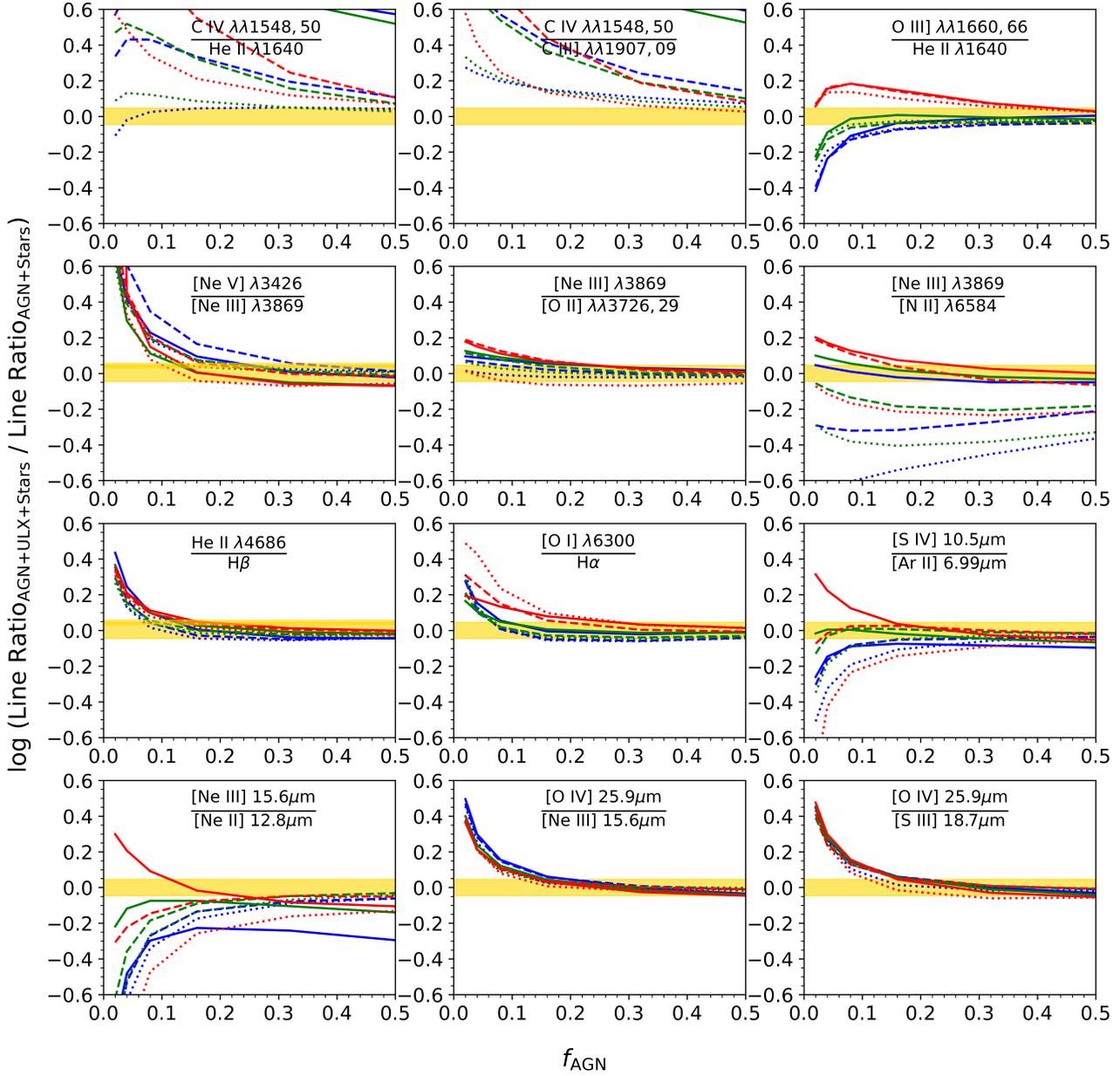

**Figure 13.** The ratio of emission line predictions assuming a post-starburst age of 25.1 Myr (i.e., maximum relative ULX contribution) to predictions assuming a post-starburst age of 1 Myr (i.e., no ULX contribution) with a light seeding channel for select AGN diagnostics in the UV-IR. The gold band represents when ULX contribution dips below 10%. The level of ULX contribution as a function of $f_{\rm AGN}$ depends upon $U$ and $Z$, without any noticeable trends according to wavelength band. Colors and linestyles are in the same format as Figure 10

### 6.2. *X-ray*

As $f_{\rm AGN}$ changes, the fraction that the IMBH and the star+ULX components contribute to the luminosity also changes. At a particular post-starburst age and $f_{\rm AGN}$ the IMBH or ULX will become the dominant contributor to X-ray emission. Figure 14 assesses when the IMBH contribution to the broadband X-ray spectrum (2-10 keV) overtakes the ULX contribution. The predicted $L_{\rm X}$ is normalized to a model where $f_{\rm AGN} = 0$ such that we can define a dashed black line in the figure where the IMBH can be considered dominant (a factor of five). The heavy seed model (top panels) shows that when the ULX first turns on (5 Myr) the IMBH remains the most influential component to the X-ray luminosity even at $f_{\rm AGN} = 0.02$. As time progresses, the IMBH signature becomes progressively more influenced by the ULX, especially at low metallicities. The heavy seeding channel has a $M_{\rm BH}$ floor of $10^5$ M$_\odot$ (Section 2), which means the black hole mass remains constant for the models in Figure 14. In contrast, the light seed model still incorporates changes in the $M_{\rm BH}$ down to log $Z/Z_\odot = -1.36$ where it floors at $M_{\rm BH} = 10^2$ M$_\odot$. This boosts the influence of the IMBH at low metallicities, making it easier to attribute X-ray emission to the IMBH at low $f_{\rm AGN}$ and higher post-starburst ages. However, since a single post-starburst age does not easily characterize dwarf galaxies due to their bursty SFHs, disentangling ULX and IMBH contributions with X-rays alone is problematic unless $f_{\rm AGN} > 0.32$, which is unlikely given the highly star-forming nature of dwarfs.

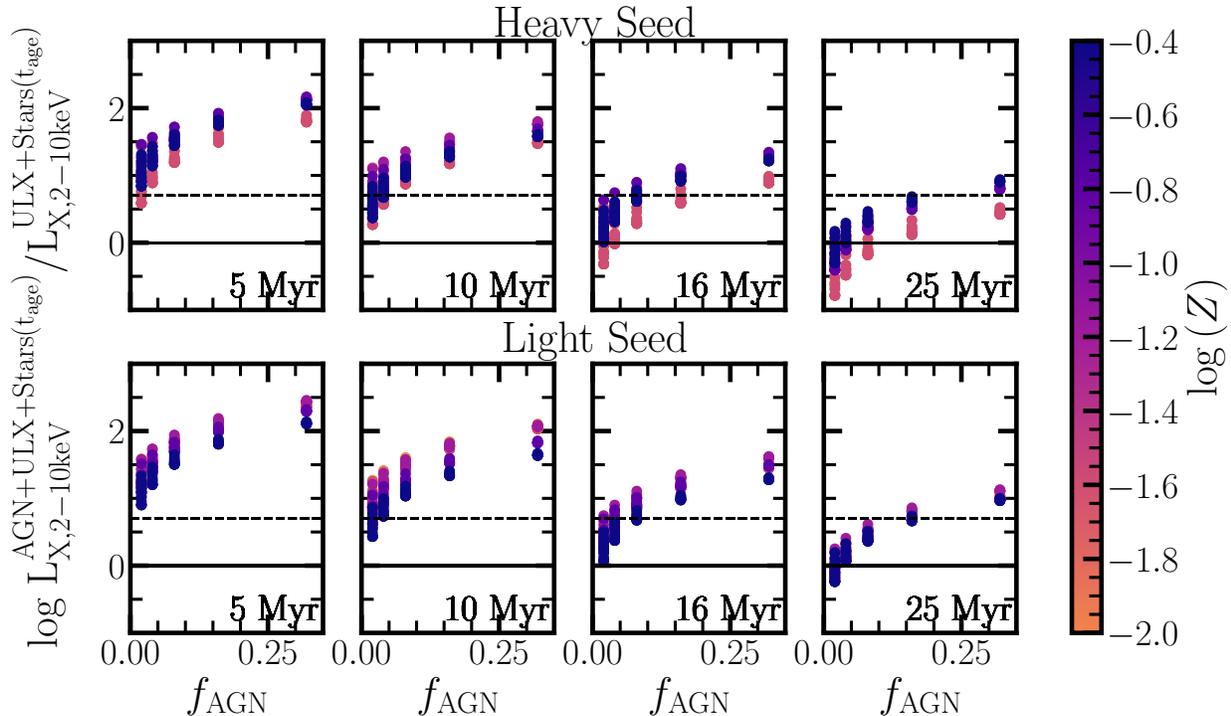

**Figure 14.** The predicted broadband (2 keV-10 keV) fluxes assuming a heavy seeding channel (top) and a light seeding channel (bottom), relative to a $f_{\rm AGN} = 0$ model. The dashed black line represents when the IMBH contribution dominates over the ULX contribution by a factor of five. The light seeding model strongly enhances $L_{\rm X}$ in the lowest metallicity simulations.

## 7. TESTING COMMON AGN CLASSIFICATION DIAGNOSTICS

Emission line ratio diagnostic diagrams provide a convenient method for distinguishing AGN since they are based on only three or four strong emission lines. However, traditional diagnostic diagrams can fail in the dwarf galaxy regime. Many of these diagrams are biased towards more massive AGN and galaxies with a higher metallicity (Cann et al. 2019; Polimera et al. 2022; Richardson et al. 2022). Dwarf galaxies are metal-poor and highly star-forming, leading to an increase in the difficulty of classification of dwarf AGN. In addition, dwarf galaxies are likely to contain a range of IMBH masses that are unaccounted for in many photoionization models (Section 4). Excluding these properties can falsely classify a dwarf AGN as a purely star-forming galaxy (Cann et al. 2019). Many of these diagnostics are also susceptible to ULX contributions (Sections 5 and 6). These ULXs can cause models to be classified as AGN using various diagnostic diagrams when, without a ULX, they would be classified as star-forming using those diagrams (Simmonds et al. 2021; Garofali et al. 2024). Diagrams have been developed to fix the lack of diagnostics for dwarfs, but a multi-wavelength systematic assessment that accounts for all the limiting factors previously mentioned is still missing. In this section, we aim to bring the effect of ULXs along with the sensitivity to $M_{\rm BH}$ together in a robust way to test the effectiveness of various emission-line diagnostic diagrams.

Figures 15-18 demonstrate the effectiveness of many diagnostic diagrams that are used to identify AGN as a function of log $Z/Z_\odot$ and $f_{\rm AGN}$ according to the demarcations listed in Tables 1-3. The color at any coordinate represents the fraction of models classified as AGN. The truth value is assessed according to our knowledge of the simulation input (i.e., $f_{\rm AGN} > 0$) relative to existing classification demarcations for AGN from the literature (listed in Tables 1-3). Dark blue corresponds to 100% of models classified as AGN by the diagnostic, and light

blue corresponds to 0% of models classified as AGN (i.e., existing demarcations do not register a dwarf AGN).

The ionization parameter range listed at the top of each subplot indicates models used to construct that particular color map based on when the line ratios used for the diagnostic diagram are bright relative to H$\alpha$ (Section 5). The post-starburst age listed at the top of each subplot corresponds to the age <5 Myr where the greatest fraction of models with $f_{\rm AGN} = 0$ would register as an AGN according to the diagnostic diagram presented in the subplot. The black dashed line on some of the diagnostics (Figures 16 and 18) represents the classifications of models when not including the composite or LINER regions for those diagnostics.

Tables 1-3 also list the "success rate" of each diagnostic diagram, which is obtained by integrating from $f_{\rm AGN} = 0.02$ to 0.5 and across all log $Z$ values. This provides a convenient metric for assessing how well each diagnostic diagram classifies dwarf AGN models as AGN. In diagrams where a composite or LINER region exists, we provide a second success rate, which excludes counting AGN models in the composite or LINER regions as AGN. The red area in each subplot indicates where each diagram would fail to robustly distinguish AGN at an age $\geq$ 5 Myr, representing possible contributions due to ULXs. We compare ULX and IMBH excitation in more detail in Section 8.

### 7.1. UV Diagrams

Figure 15 shows the parameter space where common UV diagnostic diagrams can effectively identify AGN. Figure 15a uses the diagram C III] $\lambda\lambda$1907,09 / He II $\lambda$1640 vs. O III] $\lambda\lambda$1666 / He II $\lambda$1640 first proposed by Feltre et al. (2016) with demarcations added later in Mingozzi et al. (2024). We find that the diagram is largely ineffective at identifying dwarf AGN except when close to solar metallicities as evidenced by the low success rate of 0.32. We evaluated this diagnostic assuming $-2.5 \leq \log U \leq -1$ based on the high ionization potential of He II. If we extend the ionization parameter range to lower ionization, the diagram becomes more effective yet less realistic given the high ionization potential of the lines.

Figure 15b applies the C III] $\lambda\lambda$1907,09 / He II $\lambda$1640 vs. C IV $\lambda\lambda$1548,51 / He II $\lambda$1640 diagnostic first developed using starburst and shock models in Jaskot & Ravindranath (2016) and then expanded upon with AGN models from Feltre et al. (2016) in Mingozzi et al. (2024). ULXs can cause false positive AGN detection at metallicities spanning $Z \approx [-1.25, -0.75]$, as shown by the slim red shading. Overall, however, the diagnostic is sensitive to detecting dwarf AGN over the full metallicity range of our grid once $f_{\rm AGN} \gtrsim 0.08$, translating to a success rate near 100%.

Figure 15c show the utility of C IV $\lambda\lambda$1548,51 / C III] $\lambda\lambda$1907,09 vs. (C III] $\lambda\lambda$1907,09+C IV $\lambda\lambda$1548,51) / He II $\lambda$1640 diagnostic introduced by Nakajima et al. (2018) after applying photoionization models with both stellar and AGN excitation, as well as making direct comparisons to observed SF galaxies and AGN at a variety of redshifts. The color map essentially mirrors the results from Panel (b): AGN classification is possible across a wide range of metallicities when the stellar excitation component is high (i.e., low $f_{\rm AGN}$), but ULXs can create false positive dwarf AGN detections (Section 8.1).

### 7.2. Optical Diagrams
#### 7.2.1. VO87 Diagrams

The top three subplots (a)-(c) in Figure 16 are AGN diagnostics from Veilleux & Osterbrock (1987). In particular, the BPT diagram (Baldwin et al. 1981) in Figure 16a shows [N II] $\lambda$6584/H$\alpha$ against [O III] $\lambda$5007/H$\beta$ and represents the gold standard for tracing AGN activity in giant galaxies. However, the diagnostic incorrectly identifies dwarf AGN because of their low metallicity and high star formation rates (Reines et al. 2020; Polimera et al. 2022). Figure 16a and the success rate in Table 2 corroborate this result, showing that the diagram fails to classify any models as AGN at log $Z/Z_\odot \lesssim$ -0.7 and only robustly classifies at solar metallicity with $f_{\rm AGN} > 0.16$. The diagram is also susceptible to ULX contributions for log $Z/Z_\odot \gtrsim$ -0.7. The composite region marginally affects the BPT diagram classifications as indicated by a slight decrease in the success rate from 0.23 to 0.17 in Table 2. Without this region, the diagram fails to identify most models as AGN with an $f_{\rm AGN} \leq 8\%$, and only $\sim$33% of models with $f_{\rm AGN}$ = 0.16 get classified as an AGN.

Figure 16b-c display [S II] $\lambda$6720/H$\alpha$ vs. [O III] $\lambda$5007/H$\beta$ and [O I] $\lambda$6300/H$\alpha$ vs. [O III] $\lambda$5007/H$\beta$ respectively, here on known as the [S II] diagram and [O I] diagram. These diagrams have been shown to classify dwarf AGN much better than the BPT diagram because of their metallicity insensitivity (Polimera et al. 2022, see also Reines et al. 2020). Figure 16b shows that [S II] can correctly classify a slightly greater fraction of AGN models at slightly lower metallicity (log $Z/Z_\odot \geq$ -0.9) than the BPT diagram while providing slightly more robust classifications of AGN at $f_{\rm AGN} \geq 16\%$. The [S II] diagram is susceptible to ULX contributions in the same manner as the BPT diagram; however, excluding the LINER region of this diagram (Kewley et al. 2006) is much less impactful.

Table 1. Demarcations for UV AGN Diagnostics

| Success Rate | Demarcation | Reference |
|---|---|---|
| 0.32 | $\log\left(\frac{\text{C III}]\ \lambda\lambda 1907,09}{\text{He II}\ \lambda 1640}\right) = 0.8 \log\left(\frac{[\text{O III}]\ \lambda 1666}{\text{He II}\ \lambda 1640}\right) + 0.2,\ \log\left(\frac{[\text{O III}]\ \lambda 1666}{\text{He II}\ \lambda 1640}\right) < -0.15$ | Mingozzi et al. (2024) |
| 0.98 | $\log\left(\frac{\text{C III}]\ \lambda\lambda 1907,09}{\text{He II}\ \lambda 1640}\right) = 0.43 + 0.32 \log\left(\frac{\text{C IV}\ \lambda\lambda 1548,51}{\text{He II}\ \lambda 1640}\right)$ | Jaskot & Ravindranath (2016) |
| 0.98 | $\log\left(\frac{\text{C IV}\ \lambda\lambda 1548,51}{\text{C III}]\ \lambda\lambda 1907,09}\right) = 2.5 \log\left(\frac{\text{C III}]\ \lambda\lambda 1907,09 + \text{C IV}\ \lambda\lambda 1548,51}{\text{He II}\ \lambda 1640}\right) - 1.8$ | Nakajima et al. (2018) |

NOTE—The demarcations separating AGN and SF galaxies for the UV emission line ratio diagrams presented in Figure 15. The success rate indicates the fraction of AGN models classified as AGN by each diagram over the parameter space shown in Figure 15.

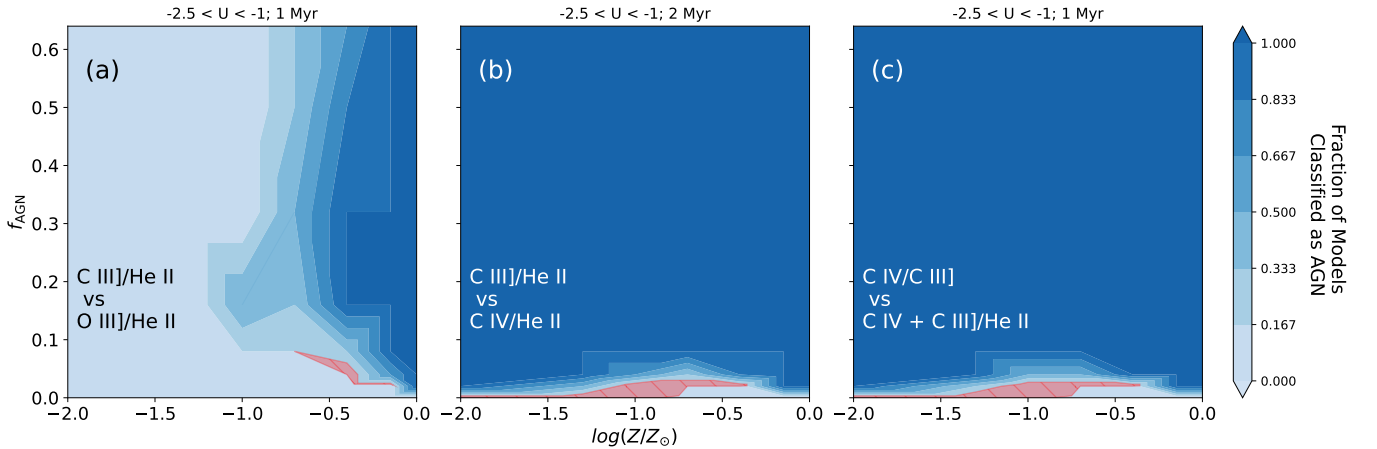

**Figure 15.** Fraction of models correctly classified as AGN with several UV diagnostic diagrams with varying physical parameters. We vary both $f_{\text{AGN}}$ and $Z$ to calculate the fraction of models that are classified as AGN as indicated by the colorbar. Each subplot uses a different diagnostic diagram to identify AGN with demarcations given in Table 1. The subplots are titled by the ionization parameter range in which they are likely to be observed as well as the age < 5 Myr at which the diagnostic diagram can categorize AGN models as AGN across the largest area of the $f_{\text{AGN}}-Z$ parameter space. The red area indicates where each diagram would distinguish AGN at an age $\geq 5$ Myr, representing false positive detection of an AGN due to a ULXs. The diagrams featuring C IV $\lambda\lambda 1548,51$ have the potential to identify dwarf AGN over a wide range of $Z$, down to low $f_{\text{AGN}}$, but ULXs can mimic the presence of AGN.

The [O I] diagram provides a significant improvement over the [S II] and BPT diagrams metal-poor environments: it classifies a greater fraction of models as AGN around $\log Z/Z_\odot \approx -0.5$ compared to the S II and BPT diagrams, which is around the typical metallicity for local dwarf AGN (Polimera et al. 2022). In general, the [O I] diagram is a strong AGN diagnostic over a broader range of metallicities compared to the [S II] diagram, which is why [O I] classified AGN are almost always classified as AGN with the [S II] diagram as well (Polimera et al. 2022). The [O I] diagram, like the other two Veilleux & Osterbrock (1987) diagrams, is susceptible to ULX contributions at low $f_{\text{AGN}}$. Excluding the LINER region (Kewley et al. 2006) of the [O I] diagram narrows the metallicity range where the diagram is the most effective down to $\log Z/Z_\odot \approx -0.75$, despite the harder IMBH radiation field (Section 4.2), as evidenced by the decrease in the success rate from 0.64 to 0.45. This aligns with results from Dors et al. (2024), which showed the low abundance of alpha elements in the interstellar medium (ISM) can cause diagnostics like [O I]/H$\alpha$ to fail at low metallicities.

### 7.2.2. Low Ionization Diagrams

Figures 16d-g feature diagrams that are most likely to be used in the ionization range of $-3.5 \leq U \leq -2.0$. Figures 16d-f both use the [O II] $\lambda\lambda$ 3726,29 emission line doublet, which has been revitalized in diagnostic diagrams in the JWST era. Figure 16d applies the [O III] $\lambda 5007$/H$\beta$ vs. [O II] $\lambda\lambda$ 3726,29/H$\beta$ diagram introduced by Lamareille (2010) that shows it only performs marginally better than the BPT diagram at identifying dwarf AGN and is affected by ULX contributions. Figure 16e displays the "OHNO" diagram

Table 2. Demarcations for Optical AGN Diagnostics

| Success Rate | Demarcation | Reference |
|---|---|---|
| | Veilleux & Osterbrock (1987) Diagrams | |
| 0.23/0.17 | $\log\left(\frac{\text{[O III] }\lambda 5007}{\text{H}\beta}\right) = \frac{0.61}{\log\left(\text{[N II] }\lambda 6584/\text{H}\alpha\right)-0.47} + 1.19$ | Kewley et al. (2001) |
| 0.34/0.30 | $\log\left(\frac{\text{[O III] }\lambda 5007}{\text{H}\beta}\right) = \frac{0.72}{\log\left(\text{[S II] }\lambda 6720/\text{H}\alpha\right)-0.32} + 1.30$ | Kewley et al. (2001) |
| 0.64/0.45 | $\log\left(\frac{\text{[O III] }\lambda 5007}{\text{H}\beta}\right) = \frac{0.73}{\log\left(\text{[O I] }\lambda 6300/\text{H}\alpha\right)+0.59} + 1.33$ | Kewley et al. (2001) |
| | Low Ionization Diagrams | |
| 0.29 | $\log\left(\frac{\text{[O III] }\lambda 5007}{\text{H}\beta}\right) = \frac{0.11}{\log\left(\text{[O II] }\lambda\lambda 3727,3729/\text{H}\beta\right)-0.92} + 0.85$ | Lamareille (2010) |
| 0.36 | $\log\left(\frac{\text{[O III] }\lambda 5007}{\text{H}\beta}\right) = \frac{0.35}{2.8\log\left(\text{[Ne III] }\lambda 3869/\text{[O II] }\lambda\lambda 3727,3729\right)-0.8} + 0.64$ | Backhaus et al. (2022) |
| 0.83 | $\log\left(\frac{\text{[O III] }\lambda 5007}{\text{[O II] }\lambda 3727}\right) = -0.84\log\left(\frac{\text{[O I] }\lambda 6300}{\text{[O III] }\lambda 5007}\right) - 1.0$ | Mingozzi et al. (2024) |
| 0.66 | $\log\left(\frac{\text{[S II] }\lambda 6720}{\text{H}\alpha}\right) = -0.48 + 0.52\log\left(\frac{\text{[N II] }\lambda 6584}{\text{[Ne III] }\lambda 3869}\right) - 0.17\left[\log\left(\frac{\text{[N II] }\lambda 6584}{\text{[Ne III] }\lambda 3869}\right)\right]^2$ | Equation 2 |
| | High Ionization Diagrams | |
| 0.83 | $\log\left(\frac{\text{[He II] }\lambda 4686}{\text{H}\beta}\right) = -1.07 + \frac{1}{8.92\log\left(\text{[N II] }\lambda 6584/\text{H}\alpha\right)-0.95}$ | Shirazi & Brinchmann (2012) |
| 1.00/0.17 | $\log\left(\frac{\text{[Ne V] }\lambda 3427}{\text{[Ne III] }\lambda 3869}\right) = -5.0$ | Cleri et al. (2023) |
| | Weak Line Diagrams | |
| 0.20 | $\log\left(\frac{\text{[O III] }\lambda 4363}{\text{H}\gamma}\right) = 0.55\log\left(\frac{\text{[O III] }\lambda 5007}{\text{[O II] }\lambda\lambda 3727,3729}\right) - 0.95$ | Mazzolari et al. (2024) |
| 0.14 | $\log\left(\frac{\text{[O III] }\lambda 4363}{\text{H}\gamma}\right) = 0.48\log\left(\frac{\text{[Ne III] }\lambda 3869}{\text{[O II] }\lambda\lambda 3727,3729}\right) - 0.42$ | Mazzolari et al. (2024) |
| 0.08 | $\log\left(\frac{\text{[O III] }\lambda 4363}{\text{H}\gamma}\right) = -1.1\log\left(\frac{\text{[O III] }\lambda 5007}{\text{[O III] }\lambda 4363}\right) + 1.47$ | Mazzolari et al. (2024) |

NOTE—The demarcations separating AGN and SF galaxies for the optical emission line ratio diagrams presented in Figure 16. The success rate indicates the fraction of AGN models classified as AGN by each diagram over the parameter space shown in Figure 16. The second number in the success rate column indicates the value if galaxies in the composite region or the LINER region of a diagram are no longer counted as AGN.

([O III] $\lambda$5007/H$\beta$ vs. [Ne III] $\lambda$ 3869/[O II] $\lambda\lambda$ 3726,29) introduced by Backhaus et al. (2022) to interpret the excitation mechanism of high-$z$ galaxies observed with JWST. The AGN region of this diagram is empirically calibrated with X-ray AGN and [Ne V] detected AGN, both uncommon for typical dwarf galaxies. Indeed, Figure 16e shows that a small fraction of AGN models are properly classified, and those models require $f_{\text{AGN}} \geq 0.32$ with relatively high metallicity.

The diagram in Figure 16f was calibrated in Mingozzi et al. (2024) using photoionization and shock models and has the virtue of only including emission lines from oxygen, [O III] $\lambda$5007/[O II] $\lambda\lambda$ 3726,39 vs. [O I] $\lambda$6300/[O III] $\lambda$5007. *Interestingly, it is the only diagnostic diagram in the UV-IR that identifies AGN at lower metallicities better than at higher metallicities.* Given the high success rate (0.83) and low probability of ULX contributions, this makes the diagram well-suited for linking dwarf AGN across a large range of metallicity when compared with other diagnostics (e.g., the [O I] diagram).

In Figure 16g we introduce a new diagram that uses [S II] $\lambda\lambda$6716,31/H$\alpha$ vs. [N II] $\lambda$6584/[Ne III] $\lambda$3869 with the following demarcation to separate AGN and SF galaxies,

$$y = -0.48 + 0.52x - 0.17x^2 \quad (2)$$

where $y = \log$ [S II]/H$\alpha$ and $x = \log$ [N II]/[Ne III]. Despite incorporating the metallicity sensitive [N II] emission line, this new diagnostic can identify dwarf AGN between $\log Z/Z_\odot$=-1.3 and $\log Z/Z_\odot$=-0.15, a much larger range than the [S II] diagram, albeit at higher AGN fractions $f_{\text{AGN}} \geq 0.32$.

#### 7.2.3. *High Ionization Diagrams*

Figures 16h and 16i contain emission lines primarily found in high ionization $-2.5 \leq \log U \leq -1.0$ galaxies. Figure 16h, proposed in Shirazi & Brinchmann (2012), uses He II $\lambda$4686/H$\beta$ vs. [N II] $\lambda$6584 to separate AGN and SF galaxies. Two demarcations are provided in Shirazi & Brinchmann (2012): an empirical demarcation based on observations and a theoretical demarcation based on photoionization models. Our tests showed that the theoretical demarcation is consistent with older versions of BPASS (v1.0), but the more up-to-date BPASS models (V2.2) are consistent with the empirical demarcation. We address these discrepancies further in Section 8. After applying the empirical dividing line for SF galaxies and AGN, Figure 16h shows that the diagnostic robustly identifies dwarf AGN across the full range of metallicities in our grid for $f_{\rm AGN} \geq 0.16$ with a high success rate of 0.83. Additionally, there are no false positive detections due to ULXs at $f_{\rm AGN} = 0$.

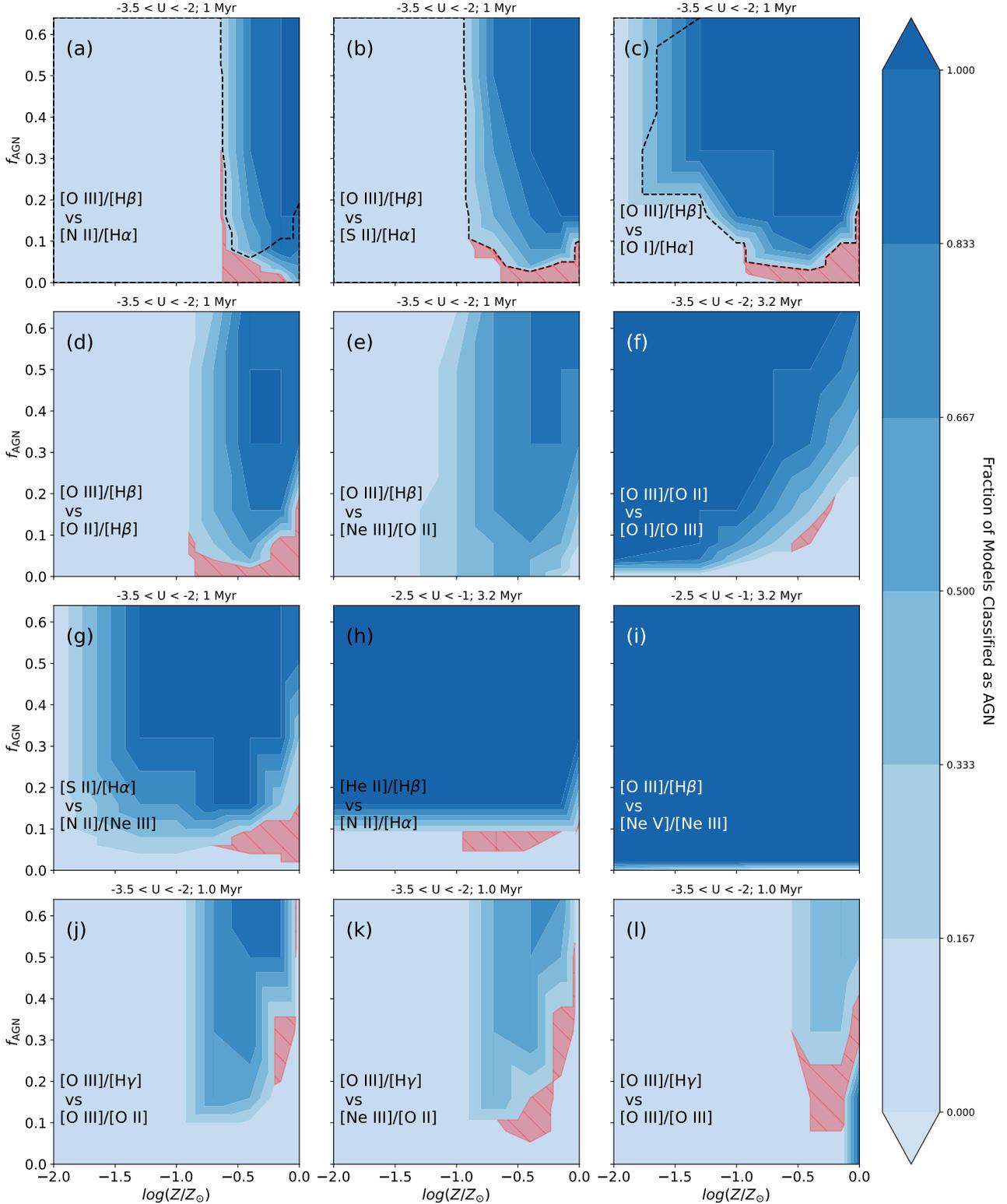

**Figure 16.** Fraction of models correctly classified as AGN with several optical diagnostic diagrams (demarcations given in Table 2) in the same manner as Figure 15. The black dashed line is shown on the subplots with diagnostic diagrams that contain a composite or LINER zone (a),(b),(c) and (i). The area contained within this line is the maximum parameter space covered when not including the composite zone for these four diagnostic diagrams. The figure shows in the Ne V diagram (Panel (i)) can distinguish AGN models for $f_{AGN}$; however, they all fall into the composite region (Figure 17). While many diagnostics can identify AGN in the metallicity range $-0.8 \leq \log Z/Z_\odot \leq -0.2$, they are susceptible to ULX contributions (red area).

Figure 16i applies the [O III] $\lambda 5007/{\rm H}\beta$ vs. [Ne V] $\lambda 3427/$[Ne III] $\lambda 3869$ diagram calibrated with photoionization models in Cleri et al. (2023). At first glance, the diagram appears immensely successful at identifying dwarf AGN at all subsolar metallicities and AGN fractions. However, Figure 17 shows that excluding the composite region of the diagram significantly narrows the parameter space in which it applies to $\log Z/Z_\odot > -0.4$ and $f_{\rm AGN} \gtrsim 0.1$, where ULXs can substantially contribute to the observed emission. The drastic contrast in the [Ne V] diagram's ability to identify dwarf AGN outside of the composite region is indicated by the sharp drop in the success rate from 1.00 to 0.17 as provided in Table 2.

### 7.2.4. [O III] λ4363 Diagrams

In the JWST era, a limited number of emission lines have caused the development of novel diagnostic diagrams involving typically weak lines. In particular, Mazzolari et al. (2024) created several diagrams based on modeling the auroral [O III] $\lambda 4363$ line (Figures 16j-l), which were explored even further in Backhaus et al. (2025). Our analysis shows that these diagrams only have a slim probability of correctly identifying dwarf AGN with success rates $\leq 0.2$. The main exception lies in the [O III] $\lambda 4363/{\rm H}\gamma$ vs. [O III] $\lambda 5007/$[O II] $\lambda 3727,29$ diagram (Figure 16j) in a narrow range of metallicities ($-0.5 \leq \log Z/Z_\odot \leq -0.15$) and $f_{\rm AGN} > 0.5$, the latter of which is uncharacteristic of dwarf galaxies with ubiquitously strong star formation.

### 7.3. IR Diagrams

Figure 18a-c displays the utility of several diagnostic diagrams using mid-IR line ratios from Richardson et al. (2022) with demarcations given in Table 3. Figure 18a implements the [Ne III] $15.6\mu m/$[Ne II] $12.8\mu m$ vs. [O IV] $25.9\mu m/$[Ne III] $15.6\mu m$ diagram first developed by Weaver et al. (2010) to separate X-ray AGN from star-forming galaxies. The demarcations were later revised in Richardson et al. (2022) using photoionization simulations, which showed that dwarf AGN would be misclassified as star-forming galaxies using the original demarcations. The diagram properly classifies AGN

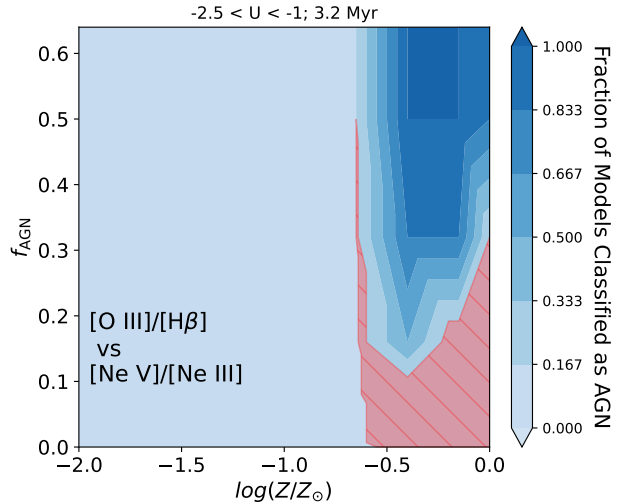

**Figure 17.** The effectiveness of the [Ne V]/[Ne III] diagnostic diagram from Figure 16 regenerated without including the composite region of that diagram. Most AGN models reside in the composite region, rendering the diagram relatively ineffective at identifying the majority dwarf AGN.

models over the entire metallicity and $f_{\rm AGN}$ range. However, ULXs can mimic AGN at $f_{\rm AGN} = 0.0$ across almost all metallicities. Excluding the composition region of the diagram provides the same success rate as Figure 18a, which highlights that the composition region in Richardson et al. (2022) was determined from observations rather than simulations. Photoionization models with only starlight excitation do not enter the composite region, which we explore in Section 8.

Figure 18b uses the [O IV] $25.9\mu m/$[Ne III] $15.6\mu m$ vs. [S IV] $10.5\mu m/$[Ne II] $12.8\mu m$ diagram developed in Richardson et al. (2022). The diagram essentially shows the same behavior as the previous diagram with [Ne III]/[Ne II] on the ordinate, including ULX contributions across virtually all metallicities.

Finally, Figure 18c displays the utility of the [O IV] $25.9\mu m/$[S III] $18.7\mu m$ vs. [S IV] $10.5\mu m/$[Ar II] $6.99\mu m$ diagram in identifying dwarf AGN. Developed in Richardson et al. (2022), the outcome is again very similar to the previous two diagrams, except the level of ULX contribution is slightly enhanced at higher metallicities.

## 8. DIAGNOSTIC DIAGRAMS FOR IDENTIFYING DWARF AGN

Ideally, as many emission lines as possible should be used to characterize the nature of the excitation source and gas conditions in a given dwarf galaxy, which we elaborate upon in 9.1.2. In practice, however, dwarf galaxies of interest may only have a handful of observed emission lines, so a deeper analysis of diagnostic dia-

Table 3. Demarcations for IR AGN Diagnostics

| Success Rate | Demarcation | Reference |
|---|---|---|
| 0.99/0.99 | $\log\left(\frac{\text{[Ne III] }15.6\mu\text{m}}{\text{[Ne II] }12.8\mu\text{m}}\right) = 0.3\log\left(\frac{\text{[O IV] }25.9\mu\text{m}}{\text{[Ne III] }15.6\mu\text{m}}\right) - 0.9$ | Richardson et al. (2022) |
| 0.99/0.99 | $\log\left(\frac{\text{[O IV] }25.9\mu\text{m}}{\text{[Ne III] }15.6\mu\text{m}}\right) = 1.25\log\left(\frac{\text{[S IV] }10.5\mu\text{m}}{\text{[Ne II] }12.8\mu\text{m}}\right) - 0.125$ | Richardson et al. (2022) |
| 1.00/1.00 | $\log\left(\frac{\text{[O IV] }25.9\mu\text{m}}{\text{[S III] }18.7\mu\text{m}}\right) = 1.2\left(\log\left(\frac{\text{[S IV] }10.5\mu\text{m}}{\text{[Ar II] }6.99\mu\text{m}}\right) + 1.4\right)^{1/2} - 3.0$ | Richardson et al. (2022) |

NOTE—The demarcations separating AGN and SF galaxies for the mid-IR emission line ratio diagrams presented in Figure 18. The success rate indicates the fraction of AGN models classified as AGN by each diagram over the parameter space shown in Figure 18.

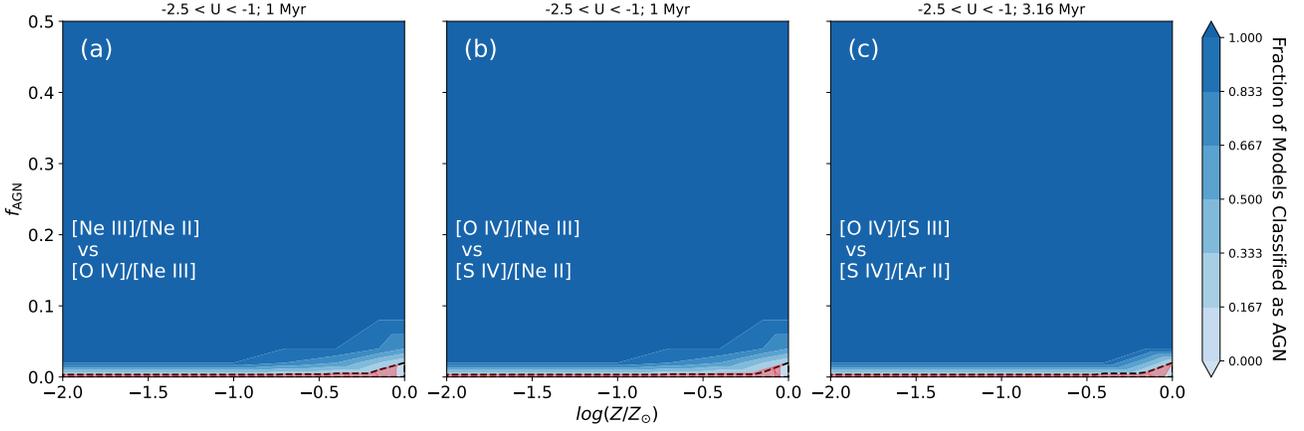

Figure 18. Fraction of models correctly classified as AGN with mid-IR diagnostic diagrams (demarcations given in Table 3) in the same manner as Figure 16. All three diagrams can identify dwarf AGN over a 2 dex metallicity range with $f_{\text{AGN}} \geq 0.02$.

grams can be very useful. The most promising diagnostic diagrams in each wavelength regime can classify dwarf AGN over a wide range of $f_{\text{AGN}}$ and metallicities with a high success rate (Figures 15-18), and detectable over a wide range of physical conditions ($f_{\text{AGN}}$, $Z$, and $U$). The second criterion rules out diagrams that use [Ne V] as a tracer (Figure 16i) since it is unlikely to be detected unless the ionization parameter and $f_{\text{AGN}}$ are large (Figure 10).

Figures 19-22 display the most promising diagnostic diagrams in each wavelength regime based on our analysis up to this point. Each figure contains several model grids. The grid labeled "BPASS" comes from the pure star-forming grid in Richardson et al. (2022) and includes the effects of binarity and WR stars without a ULX. For the "ULX+BPASS" models, we use the models developed in this paper while setting $f_{\text{AGN}} = 0$ and the post-starburst age to 25.1 Myr, maximizing the effects of the ULX relative to stars. To include active IMBHs without ULXs, we set the post-starburst age to 1 Myr and select the light seeding channel for the models labeled "AGN+BPASS". The "AGN+ULX+BPASS" grids assume a post-starburst age of 25.1 Myr and a light seeding channel. In all the grids that include active IMBHs, we restrict the AGN fraction to $0.02 \leq f_{\text{AGN}} \leq 0.5$. We limit the model space for the comparison BPASS grids to represent the max SF case, and likewise restrict our models (BPASS + AGN and BPASS + ULX + AGN) to ionization parameters and metallicities where lines are most likely detectable and will result in the strongest line signatures.

To explore other possible excitation mechanisms, we include shock models from Alarie & Morisset (2019) with the abundance scaling from Gutkin et al. (2016) assuming (C/O)/[(C/O)$_\odot$]=1. For our fiducial shock models, we selected metallicities ranging from $Z/Z_\odot = 0.01$ to $Z = Z_\odot$ and fast shocks with $v_{\text{shock}} > 100$ km s$^{-1}$. We assumed a pre-shock density of $n_{\text{H}} = 1$ cm$^{-3}$ and a magnetic field of $B = 5$ $\mu$G since varying these parameters affect the line ratios less than $v_{\text{shock}}$ and $Z$, however we tested the full range of these parameters as well.

### 8.1. UV

Figure 19 shows the C III] $\lambda\lambda$1907,09 / He II $\lambda$1640 vs. C IV $\lambda\lambda$1548,51 / He II $\lambda$1640 diagnostic diagram in the first row and the C IV $\lambda\lambda$1548,51 / C III] $\lambda\lambda$1907,09 vs. (C III] $\lambda\lambda$1907,09+C IV $\lambda\lambda$1548,51) / He II $\lambda$1640 diagnostic diagram in the second row. We restrict all the model grids to $-2.5 \leq \log U \leq -1$. The BPASS, BPASS+ULX, and shock grids are limited to all available metallicities below solar. The BPASS grid uses a post-starburst age of 250 Myr as this generated line ratios that came closest to the AGN demarcation. The BPASS+AGN and BPASS+ULX+AGN are shown for $Z = 0.1\ Z_\odot$ since the AGN classification for both diagrams is relatively independent of metallicity (Fig 15).

Both diagrams cleanly separate star-forming galaxies from dwarf AGN as indicated in Figures 15b and c. However, Figure 19 shows that ULXs have a strong potential to generate false positive AGN classifications. The BPASS+AGN and BPASS+ULX+AGN grids essentially overlap with the BPASS+ULX grid, making distinguishing between the excitation mechanisms complicated.

### 8.2. Optical

The optical presents much more promise in separating ULX and AGN excitation. Notably, the diagrams selected in the section allow us to define *new demarcations* where these already promising diagnostics are even more robust for dwarf AGN identification. In particular, we can define an "AGN threshold" beyond which our models suggest *at least some AGN excitation* is required to explain the line ratios; furthermore, we can define regions free of substantial influence from ULXs (i.e., $f_{\rm AGN}$ at and above which the AGN dominates the ionizing spectrum). Effectively, we are defining new "composite" regions (some AGN, some stars + ULX) and pure "dwarf AGN" regimes.

The [O I] $\lambda$6300 emission line is a strong diagnostic for identifying and characterizing dwarf AGN. Figure 20 displays two emission-line diagnostic diagrams featuring this line: [O III]/H$\beta$ vs. [O I]/H$\alpha$ (top panels) and [O III]/[O II] vs. [O I]/[O III] (bottom panels). The displayed models grids are in the range $-3.5 \leq \log U \leq -2.0$, which is where [O I] is the strongest (Figure 11). The BPASS, BPASS+ULX, and shock grids include all subsolar metallicities. The BPASS grid and BPASS+ULX grids use post-starburst ages of 20 Myr and 15.8 Myr, respectively, since these simulations came the closest to the AGN demarcations.

For the top panels of Figure 20, the BPASS+AGN and BPASS+ULX+AGN grids only include $Z = 0.7\ Z_\odot$ while in the bottom panels, these grids only include $Z = 0.01\ Z_\odot$, because these specific metallicities result in the greatest [O I] emission that extends past the SF/AGN demarcations (Kewley et al. 2001; Mingozzi et al. 2024). Simulations that include only binary stars (BPASS) can extend slightly past the maximum starburst demarcations in the top panels. The BPASS+ULX grids surpass the maximum starburst demarcation by a non-negligible amount in both diagrams. However, unlike in the diagnostics in UV, the BPASS+ULX grid and AGN grids eventually separate from one another. This allows an "AGN threshold" to be defined, which indicates where AGN excitation is likely required to reproduce the observed emission. For the [O III]/H$\beta$ vs. [O I]/H$\alpha$ diagram, this corresponds to

$$y = 0.48/(x + 0.59) + 1.4 \quad (3)$$

where $y =$ [O III]/H$\beta$ and $x =$ [O I]/H$\alpha$. For the [O III]/[O II] vs. [O I]/[O III] diagram, this corresponds to

$$y = -1.4x - 1.2 \quad (4)$$

where $y =$ [O III]/[O II] and $x =$ [O I]/[O III].

Additionally, the ULX continues to contribute to the BPASS+ULX+AGN grid until $f_{\rm AGN} = 0.5$, in agreement with Figure 13, after which the AGN completely dominates the ionizing spectrum. For the [O III]/H$\beta$ vs. [O I]/H$\alpha$ diagram, the approximate boundary where this occurs is defined by

$$y = 0.4/(x + 0.3) + 1.7 \quad (5)$$

while in the [O III]/[O II] vs. [O I]/[O III] diagram this is defined by

$$y = -1.4x \quad (6)$$

The shock excitation grids mostly overlap with the LINER region of the [O I]/H$\alpha$ diagnostic diagram as defined in Kewley et al. (2006) even with extended parameter values of $B$, $n_{\rm H}$, and $v_{\rm shock}$. In the bottom panels of Figure 20, the shock grids show far less overlap with those that include AGN. Overall, the results indicate both diagrams with [O I] are a reasonable tool for assessing active IMBHs in dwarf galaxies as long as the $f_{\rm AGN} \gtrsim 0.16$ enough to rule out ULXs. The two diagrams are complementary since [O III]/H$\beta$ vs. [O I]/H$\alpha$ is better suited for classifying dwarf at intermediate metallicities (Figure 16c) and the [O III]/[O II] vs. [O I]/[O III] is better suited for classifying AGN in the extremely metal-poor regime (Figure 16f).

The top panels of Figure 21 display He II/H$\beta$ vs. [N II]/H$\alpha$, a diagram introduced by Shirazi & Brinchmann (2012). The [N II]/H$\alpha$ ratio is primarily sensitive to metallicity due to secondary nucleosynthesis, while strong He II is typically seen in galaxies best fit by high ionization parameter models given the >54 eV required to produce it. For these reasons, we select subsolar metallicities for all model grids and an ionization parameter of $\log U = -2.0$ for all grids except shocks. We selected a post-starburst age of 250 Myr for the BPASS grid since this maximized the He II/H$\beta$ ratio. Additionally, we add a purple-shaded region to indicate the approximate parameter space spanned by the GALSEVN (Lecroq et al. 2024) models, which include binary star evolution.

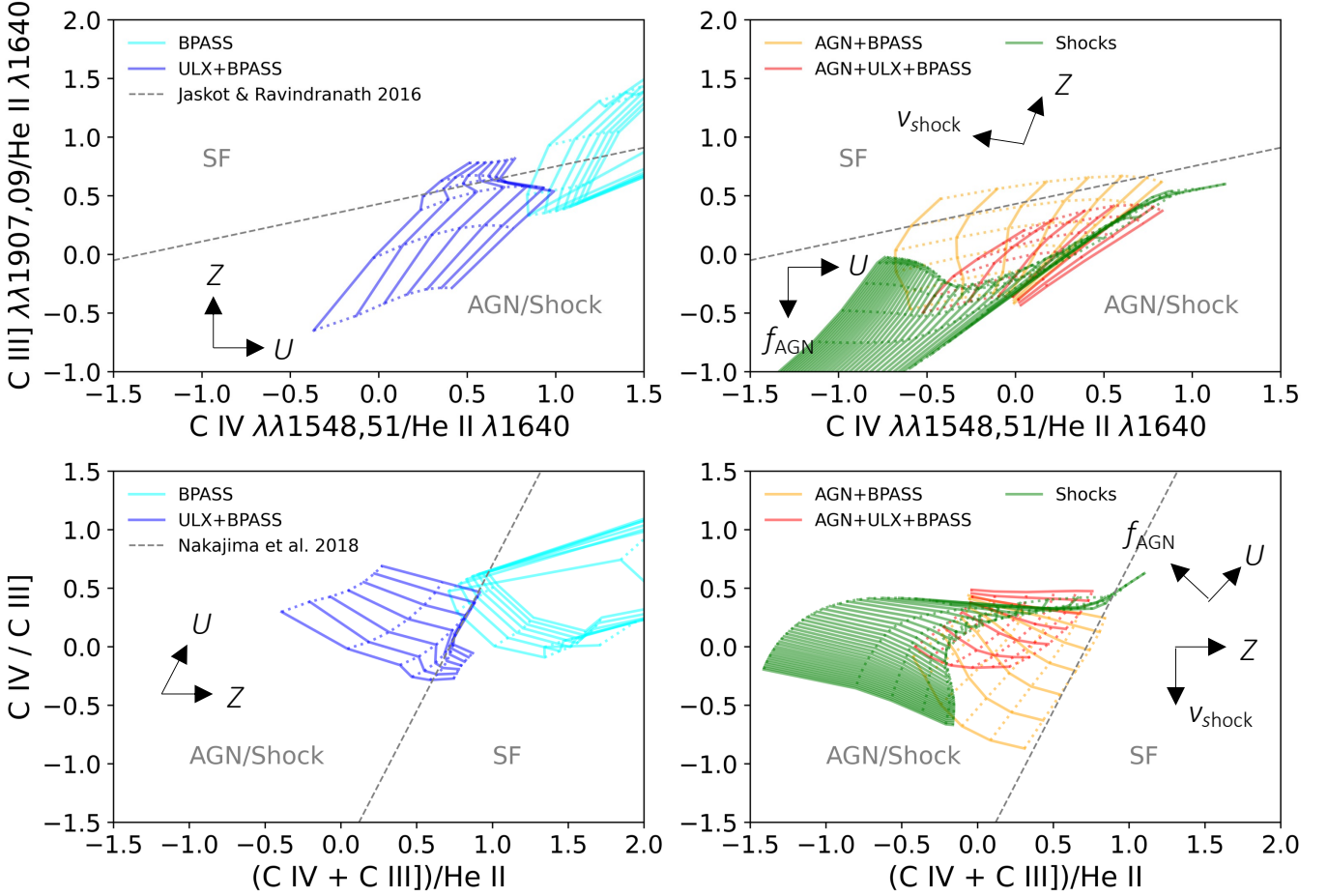

**Figure 19.** UV diagnostic diagrams capable of cleanly separating excitation from active IMBHs and starlight. The BPASS+ULX grid is capable of mimicking dwarf AGN emission to a significant degree, which strongly hinders the ability to robustly identify dwarf AGN in the UV.

The GALSEVN models provide a much harder radiation field than the BPASS models, essentially matching the BPASS+ULX grids in He II/H$\beta$ output. Unlike [O I] diagrams in Figure 21, the He II diagram shows much greater separation between the AGN grids and BPASS+ULX grid, and the IMBH SED dominates the ULX at a lower AGN fraction, $f_{\rm AGN} = 0.16$. As in the previous figure, we determined an AGN threshold demarcation given by $\log {\rm He\,II}/{\rm H}\beta = -1.6$, and the approximate boundary where the AGN dominates the ionizing spectrum is given by

$$y = 0.073x - 1.0 \qquad (7)$$

where $y = \log {\rm He\,II}/{\rm H}\beta$ and $x = \log {\rm [N\,II]}/{\rm H}\alpha$.

The shock grids overlap with highest $f_{\rm AGN}$ models, but show an offset of 0.5 dex with [N II]/H$\alpha$. The latter effect could be due to different abundances prescriptions between the shock and AGN grids, or $v_{\rm shock}$ could uniformly enhance [N II]/H$\alpha$, which does not occur with $f_{\rm AGN}$.

The bottom panels of Figure 21 display [S II]/H$\alpha$ vs. [N II]/[Ne III], a new diagnostic introduced in this paper (Section 7.2). We select an ionization parameter range $-3.5 \leq \log U \leq -2.0$ for the displayed grids given the lower ionization potentials involved in the line ratios. A post-starburst age of 20 Myr showed the greatest potential overlap with the AGN grids therefore we adopted that age for the BPASS grid. We used all the available subsolar metallicities for the BPASS, BPASS+ULX, and shock grids. We chose to display the $Z = 0.4\,Z_\odot$ models for the grids involving AGN since it represents the metallicity midpoint for dwarf AGN detectability according to Figure 16.

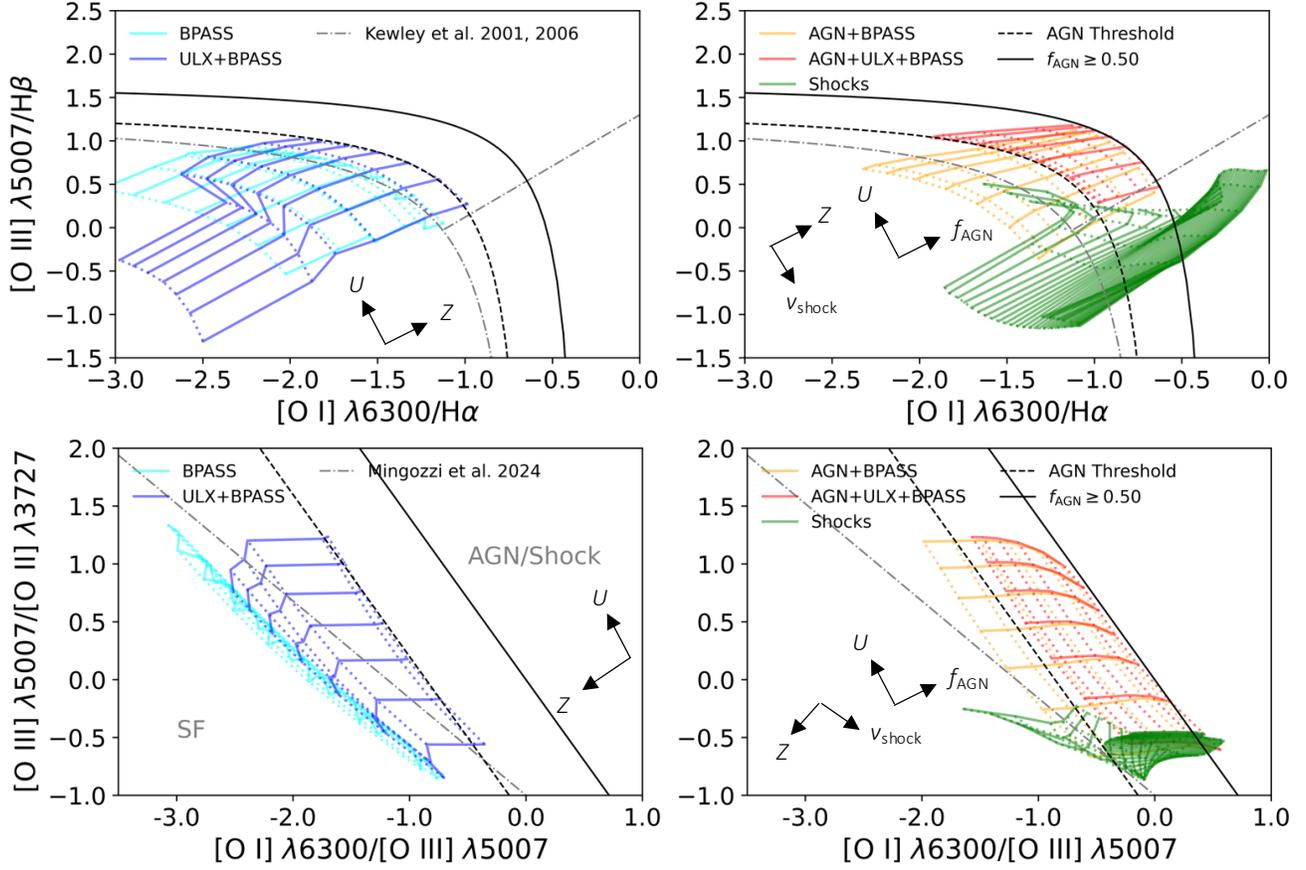

**Figure 20.** Optical diagnostic diagrams using [O I] λ6300 show promise in separating active IMBHs and starlight excitation while minimizing ULX contributions. Unlike UV diagnostics, the BPASS+ULX model grids do not overlap with the AGN grids in either diagram for $f_{\rm AGN} \gtrsim 0.16$, which defines each diagram's "AGN Threshold" demarcations. Together, the diagrams enable the identification of dwarf AGN spanning $\log Z/Z_\odot =$ -2.0 to $\log Z/Z_\odot =$ 0.0.

The BPASS and BPASS+ULX grids are cleanly separated from the AGN grids using the demarcation given by equation 2. Models with $f_{\rm AGN} >$0.16 lie outside the region dominated by stars and ULXs. However, the shock models strongly overlap with many of the AGN grids, possibly making it problematic to disentangle the two excitation mechanisms. The AGN dominates the spectrum at $f_{\rm AGN} \approx 0.5$ given by demarcation,

$$y = 0.60x - 0.18 \qquad (8)$$

where $y = \log$ [S II]/H$\alpha$ and $x = \log$ [N II]/[Ne III].

### 8.3. IR

Figure 22 displays three separate mid-IR diagnostic diagrams from Richardson et al. (2022). In these diagrams, we select an ionization parameter range of $-2.5 \leq \log U \leq -1.0$ due to the high ionization potential of the [O IV] 25.9 $\mu$m line featured in each diagram. A post-starburst age of 250 Myr comes the closest to overlapping with the AGN grids, so we select this for the stars-only model. Figure 18 shows that AGN classification is independent of metallicity in all three diagrams, so we display the $Z = 0.05$ Z$_\odot$ models for all the AGN grids as we have yet to probe this regime in the other diagnostic diagram we have analyzed.

The upper left panel of Figure 22 displays the diagram first developed in Weaver et al. (2010) to separate AGN and SF galaxies. In Richardson et al. (2022), the demarcations were redefined to include regions with dwarf AGN. The figure shows that SF excitation is cleanly separated from all other excitation mechanisms, primarily because of the [O IV] 25.9 $\mu$m line. Unlike in Richardson et al. (2022), we can define a new AGN threshold demarcation that separates stars+ULX excitation from excitation that requires an AGN,

$$y = 0.5x - 0.1 \qquad (9)$$

The location of this demarcation is different from the one that would have been created using the Garofali et al. (2024) ULX models for two reasons: (1) the ULX models in Garofali et al. (2024) stopped when the gas electron temperature dropped below 4000 K, while our simulations stopped when the electron density reached 10% of the hydrogen density, resulting in some of our simulations proceeding deeper into the cloud where low ionization species are excited; (2) our models extend down to 0.01 $Z_\odot$ while the Garofali et al. (2024) models stopped at 0.05 $Z_\odot$. Together, these differences change the progression of the [Ne III]/[Ne II] ratio for the ULX grids.

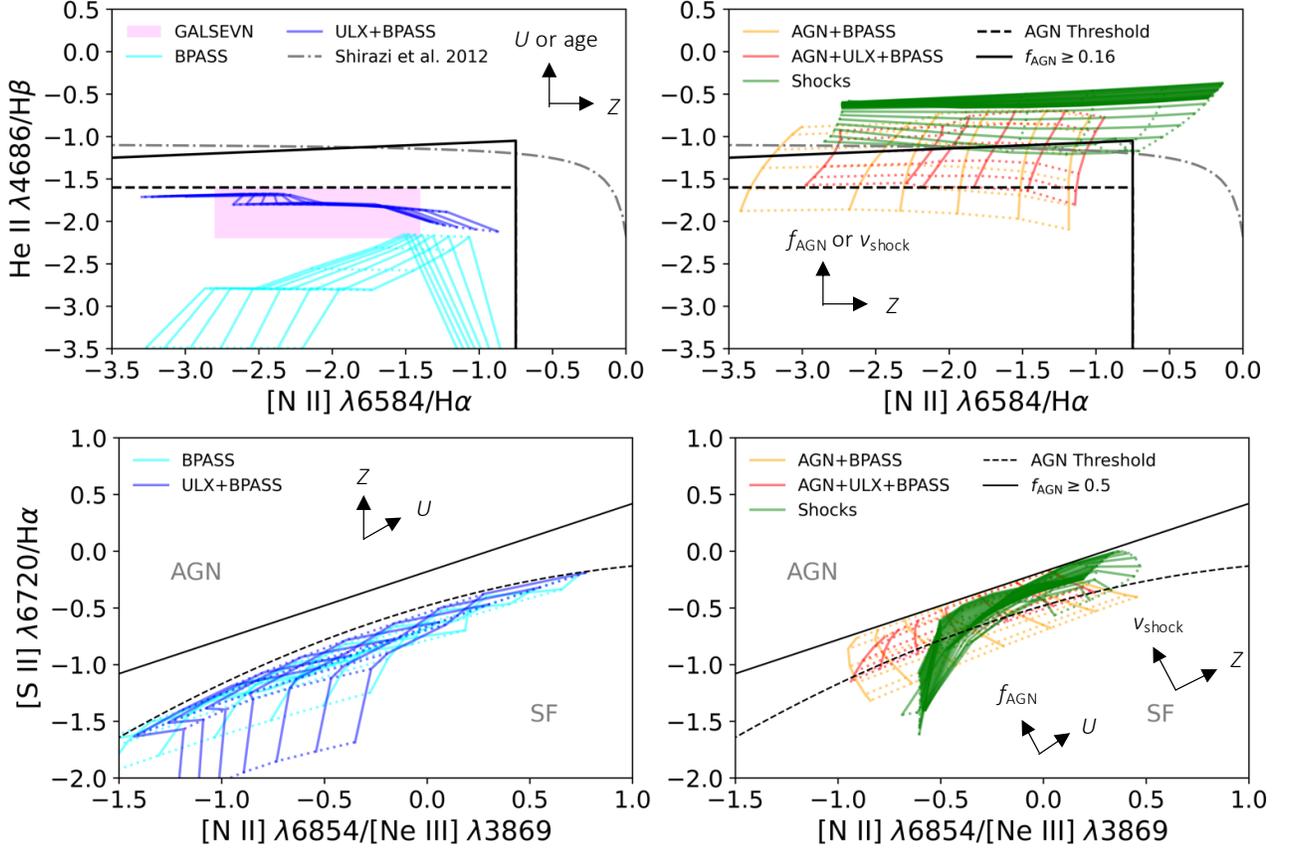

**Figure 21.** Optical diagnostic diagrams that show promise in separating active IMBHs and starlight excitation while minimizing ULX contributions. Unlike UV diagnostics, the BPASS+ULX model grids have no overlap with the AGN grids in the He II diagram for $f_{\rm AGN} \gtrsim 0.04$ and no overlap with the AGN grids in the [Ne III] diagram for $f_{\rm AGN} \gtrsim 0.16$. These limits are defined as "AGN Threshold" demarcations in each diagram.

Shock excitation does overlap the AGN grids at high AGN fractions. However, in the remaining panels of Figure 22, the shock grid is cleanly separated from all other excitation mechanisms, mainly on account of [Ne II] 12.8 $\mu$m and [Ar II] 6.99 $\mu$m emission lines forming close to the ionization front. Similarly, there is a greater separation between the ULX and AGN grids in these two diagrams, with the ULX only overlapping with models that have $f_{\rm AGN} < 0.08$. For the [S IV]/[Ne II] diagram, we define the AGN threshold according to the equation,

$$y = 0.87x - 0.087 \quad (10)$$

and for the [S IV]/[Ar II] diagram we define the AGN threshold demarcation as,

$$y = 1.09x - 0.9 \quad (11)$$

In aggregate, the three emission line ratio diagrams in Figure 22 relatively cleanly separate stellar, ULX, AGN, and shock excitation.

## 9. DISCUSSION

### 9.1. Best Practices Characterizing IMBHs in Dwarf AGN

#### 9.1.1. Specific Line Ratios

As mentioned in Section 1, previous studies have used different emission line diagnostics to characterize IMBHs in dwarf AGN, spanning the X-rays to IR. In this paper, we have shown that several line diagnostics fail to identify AGN in the metal-poor, highly star-forming regime (see also Sartori et al. 2015; Polimera et al. 2022), and many line diagnostics are affected by possible ULX contributions (see Garofali et al. 2024). In fact, *no single diagnostic line ratio is foolproof for identifying, or characterizing, dwarf AGN.* Each emission line ratio has unique limitations regarding detectability, inferring seeding and $f_{AGN}$, and distinguishing excitation mechanisms (e.g., ULXs, shocks), which state-of-art statistical tools have just started to disentangle (Einig et al. 2024). In Table 4, we have provided a qualitative summary of the more useful emission line ratios in the UV through IR based on our analysis in Sections 4-8. The emission line ratios are separated into two tiers, with the top tier representing ratios that have the best combination of diagnostic value and detectability.

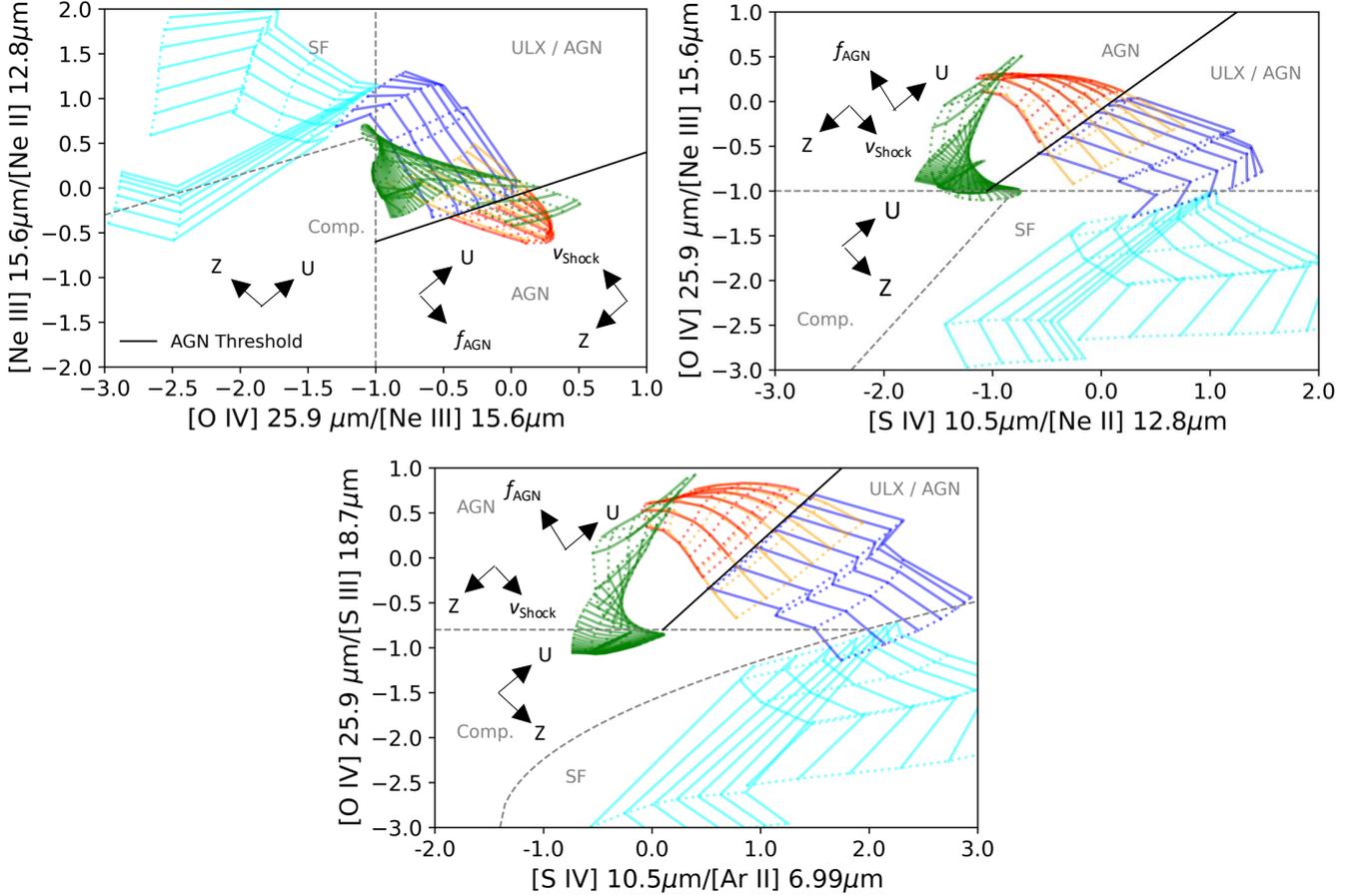

**Figure 22.** Mid-IR diagnostic diagrams showing the most promise in separating active IMBHs and starlight excitation while minimizing possible ULX contributions. The models shown represent excitation due to stars (cyan), stars+ULX (blue), stars+AGN (orange), stars+ULX+AGN (red), and shocks (green). Unlike the UV and optical diagnostics, the [O IV]/[Ne III] vs. [S IV]/[Ne II] and [O IV]/[S III] vs. [S IV]/[Ar II] diagrams show a relatively clean distinction between excitation due to stars, ULXs, AGN, and shocks.

In the top tier, He II/H$\beta$ is compelling for identifying dwarf AGN and inferring their properties. The line ratio is moderately sensitive to seeding and $M_{BH}$ with minimal overlap between AGN grids and stellar and ULX grids at moderate to high ionization parameters across a wide range of metallicities. The pitfalls of using He II $\lambda$ are twofold: (1) WR atmospheres can produce He II $\lambda 4686$ (Shirazi & Brinchmann 2012); (2) shock excitation can easily reproduce He II/H$\beta$ values similar to AGN.

The line ratio [O I]/H$\alpha$ is sensitive to seeding and $M_{BH}$, with moderate overlap between AGN grids and stellar, ULX, and shock grids. Despite the possible overlap with excitation mechanisms other than AGN, [O I] $\lambda 6300$ is uniquely detectable in low to moderate ionization gas with a low AGN fraction, unlike other optical or UV lines listed in Table 4. *Since most dwarfs are low-ionization and highly star-forming, this makes [O I] $\lambda 6300$ one the most useful line diagnostics in the optical for finding and characterizing dwarf AGN.*

Many challenges at other wavelengths for finding and characterizing AGN are mitigated at mid-IR wave-

Table 4. Most Useful Emission Line Ratios for Identifying and Characterizing Dwarf AGN

| Line Ratio | $M_{\rm BH}$ Constraint | $f_{\rm AGN}$ Constraint | Stars & AGN Separation | ULX & AGN Separation | Shock & AGN Separation | Best Conditions for Detection |
|---|---|---|---|---|---|---|
| Top Tier Diagnostics ||||||| 
| $\frac{\text{He II } \lambda 4686}{\text{H}\alpha}$ | Moderate | Moderate | Strong | Strong | Weak | Mid - High $U$ |
| $\frac{\text{[O I] } \lambda 6300}{\text{H}\alpha}$ | Strong | Moderate | Moderate | Moderate | Moderate | Low - Mid $U$ |
| $\frac{\text{[S IV] } 10.5\ \mu\text{m}}{\text{[Ar II] } 6.99\ \mu\text{m}}$ | Strong | Strong | Weak | Strong | Strong | Mid - High $U$ |
| $\frac{\text{[S IV] } 10.5\ \mu\text{m}}{\text{[Ne II] } 12.8\ \mu\text{m}}$ | Strong | Strong | Weak | Strong | Strong | Mid - High $U$ |
| $\frac{\text{[Ne III] } 15.6\ \mu\text{m}}{\text{[Ne II] } 12.8\ \mu\text{m}}$ | Strong | Strong | Moderate | Moderate | Weak | Mid - High $U$ |
| $\frac{\text{[O IV] } 25.9\ \mu\text{m}}{\text{[Ne III] } 15.6\ \mu\text{m}}$ | Strong | Strong | Strong | Weak | Strong | Mid - High $U$ |
| Second Tier Diagnostics ||||||| 
| $\frac{\text{C IV + C III]}}{\text{He II } \lambda 1640}$ | Weak | Weak | Strong | Weak | Strong | Mid - High $U$ |
| $\frac{\text{[Ne V] } \lambda 3426}{\text{[Ne III]} \lambda 3869}$ | Strong | Strong | Strong | Strong | Strong | High $U$ |
| $\frac{\text{[S II] } \lambda 6720}{\text{H}\alpha}$ | Strong | Moderate | Moderate | Moderate | Strong | Low - High $U$ |
| $\frac{\text{[Ne V] } 14.3\ \mu\text{m}}{\text{[Ne III] } 15.6\ \mu\text{m}}$ | Strong | Strong | Strong | Strong | Strong | High $U$ |

Note—For each line ratio, the sensitivity to $M_{\rm BH}$ (and therefore seeding channel), and $f_{\rm AGN}$ are given in qualitative terms. This is followed by the degree to which AGN activity can be separated from other excitation mechanisms. Finally, we list the ionization state of the gas where all the lines in a given ratio are typically detectable. Line ratios are separated into two tiers, with the top tier ratios presenting the best combination of diagnostic value and detectability.

lengths. Essentially all the listed mid-IR line ratios in Table 4 strongly trace seeding mechanisms and black hole mass. The line ratio [S IV]/[Ar II] works well in the moderate to high ionization regime across a range AGN fractions without substantial shock contamination. Yet, excitation from stars and ULXs can mimic AGN activity. The [Ne III]/[Ne II] line ratio is commonly observed over a wide range of ionization and moderately distinguishes AGN excitation from ULX and stellar sources, but shocks can often reproduce the same line ratios as AGN. Finally, [O IV]/[Ne III] strongly separates the AGN grids from stars and shock grids in the mid- to high-ionization range where the [O IV] 25.9 $\mu$m line is detectable, however ULXs can create the same emission line ratios as AGN.

Emission line ratios in the second tier of Table 4 either strongly overlap with ULXs, have limited potential in characterizing IMBH properties (e.g., $M_{\rm BH}$), or are not easily detected in most dwarfs. In the UV, the emission lines we have investigated show the least sensitivity to $M_{\rm BH}$ and seeding, while displaying a strong potential for false positive AGN classification due to ULXs. However,

C IV/He II and C IV+C III/He II can be used with other line ratios to separate stellar excitation sources from AGN and shocks at moderate to high ionization parameters (Figure 19). It is important to note that the He II $\lambda$1640 line used in these ratios can be generated in the atmospheres of WR stars, which can be confused as nebular in origin (Shirazi & Brinchmann 2012).

Spanning the blue side of the optical, the [Ne V]/[Ne III] ratio is very sensitive to seeding channel and $M_{\rm BH}$, and shows minimal overlap with stellar, AGN, ULX, and shock grids. Despite these advantages, the main drawback of the [Ne V] $\lambda$3426 line is that it weakly emits in most dwarf AGN except in high ionization, high $f_{\rm AGN}$ conditions (Chisholm et al. 2024), unlike those found in local dwarfs. Similarly, the mid-IR [Ne V] 14.3 $\mu$m line is typically a robust indicator of AGN activity, but it still requires high ionization cloud conditions with a low AGN fraction. *When detected in dwarfs, [Ne V] is usually a robust indicator of IMBH activity, however a non-detection does not exclude the possible of an active IMBH strongly influencing the emission line spectrum.*

The [S II]/H$\alpha$ ratio generally provides the same virtues of [O I]/H$\alpha$ but to a lesser extent. It shows slightly less variation with $M_{\rm BH}$ seeding, and $f_{\rm AGN}$ compared to [O I]/H$\alpha$ with the same amount of overlap with other grids. While [S II]/H$\alpha$ has less diagnostic value than [O I]/H$\alpha$, it is easier to detect in most galaxies, making it a viable substitute when [O I] and He II are unavailable.

#### 9.1.2. *Applying IMBH models to observation*

A dwarf galaxy classified as purely star-forming in an emission line ratio diagram could still have an active IMBH since each diagram is subject to bias. When many emission lines are observed, testing every possible diagnostic diagram is impractical and can give conflicting results. While diagnostic diagrams present a convenient means for identifying dwarf AGN when only a handful of emission lines are observed, none are entirely foolproof, and the diagrams themselves do not characterize the properties of the AGN (e.g., $M_{\rm BH}$, $f_{\rm AGN}$), which are arguably more meaningful than finding the AGN itself. Instead of asking whether a dwarf galaxy contains an AGN, we should ask: *how much AGN activity is present?* Determining the AGN fraction in dwarfs is not just a matter of using emission lines with widely separated ionization potentials – one must also consider the excitation energies (Figure 9) and line detectability (Section 5). This is particularly relevant for high-$z$ studies with JWST (which often probe rest optical or UV) and local studies with JWST.

To characterize the properties of the IMBH, we strongly suggest a probabilistic approach that uses all available emission lines to infer the AGN fraction, seeding mechanism, etc. Such an approach requires models complex enough to replicate the conditions within the dwarf galaxy, which might involve different combinations of excitation mechanisms. This paper has taken one step in that direction by combining ULX and AGN sources in a single model. One could imagine even more complex conditions with shocks mixed into the excitation. Additionally, the clouds surrounding the sources could take on statistical distributions of different properties, featuring a wide range of ionization states, nebular densities, etc. The complexity of these models becomes unwieldy very quickly, so novel approaches have been developed to take the problem on the fly by making the combination of excitation sources and nebular properties a free parameter in Bayesian inference (e.g., MULTIGRIS; Lebouteiller & Ramambason 2022; Ramambason et al. 2022, 2024; Morisset et al. 2025). While this analysis can be more time-consuming than quickly using strong-line calibrations or diagnostic diagrams, the results are more robust and often contradict the results from simpler approaches (Lebouteiller et al. 2025; Morisset et al. 2025).

#### 9.2. *Interpreting Recent Searches for Dwarf AGN*

In the JWST-era, high-$z$ ($z \gtrsim 5$) AGN with rest-frame UV / optical spectroscopy have generated significant interest due to their prevalence and overly massive BHs relative to local scaling relations (Maiolino et al. 2023; Greene et al. 2024). Identifying these AGN has primarily revolved around using broadened emission lines, or [Ne III] $\lambda$3869 and [Ne V] $\lambda$3426 on the blue side of the optical (Backhaus et al. 2022; Cleri et al. 2023; Chisholm et al. 2024). However, as shown in Figure 16e, the "OHNO Diagram" (Backhaus et al. 2022) cannot reliably classify dwarf AGN unless the $f_{\rm AGN} > 0.5$, which is atypical of dwarf AGN in the local universe (Polimera et al. 2022). Similarly, while [Ne V]/[Ne III] strongly traces $f_{\rm AGN}$ and $M_{\rm BH}$, our models predict that dwarf AGN with $f_{\rm AGN} < 0.5$ would all fall within the composite region of the [Ne V]/[Ne III] diagram (Figure 17) leading to very low success rate (Table 2, and [Ne V] would be difficult to detect (Figure 10). Indeed, the IMBH source in Chisholm et al. (2024) lies in the composite region of the [Ne V]/[Ne III] diagram even though the broadened [Ne V] emission is best reproduced with a model assuming $f_{\rm AGN} \approx 0.32$.

Within the context of these JWST observations, our results imply two key points: (1) the fraction of dwarf galaxies hosting AGN at high-$z$ could be even higher than current estimates because the diagnostics used to identify them require vastly different physical conditions (high $U$, $f_{\rm AGN}$, and $L_{bol}$) than those found at $z = 0$; (2) Since $U \propto L_{bol} \propto M_{BH}$, it is perhaps unsurprising that the samples of AGN detected with JWST at high-$z$ have relatively overmassive BHs embedded in locally uncommon gas conditions. These overmassive BHs at cosmic dawn have led many to postulate that heavy seeding was necessary to form them, however it is unclear if heavy seeding was the *predominant* mechanism for forming the first BHs (Regan & Volonteri 2024).

Identifying less bolometrically luminous AGN at very high-$z$ could prove problematic since the UV-rest frame emission lines picked up by NIRspec are relatively insensitive to $M_{\rm BH}$ (Figure 6) and subject to ULX contributions (Figure 19). As we noted previously, however, instead of using a handful of emission lines in various diagrams, using *all* the available emission lines together in a Bayesian framework to assess the *probability* of an AGN being present would more reliably test the fraction of dwarf galaxies with AGN at high-$z$.

It is possible that the best approach to constraining the dwarf AGN distribution could lie at $z \sim 0$, searching for relatively unevolved IMBHs in dwarf galaxies with more constraining optical and mid-IR observations. At $z \sim 0$, coronal lines were the subject of much interest in the pre-JWST era (Cann et al. 2018; Satyapal et al. 2021) on account of these high-ionization potential tracing the high-temperature accretion disk characteristic of low IMBHs. Yet, in the JWST era, dwarf AGN lack ubiquitous coronal emission line emission in the optical and mid-IR (Molina et al. 2021; Salehirad et al. 2022; Negus et al. 2023; Doan et al. 2024). Indeed, our simulations with $f_{\rm AGN} < 0.5$ do not produce detectable coronal line emission (e.g., [Si VI] 1.96$\mu$m), and unrealistically high values of $f_{\rm AGN} \approx 1$ produce modest coronal line emission even at high $U$. This suggests that the highly star-forming nature of dwarf galaxies expected to host these IMBHs strongly masks coronal line signatures. While several free parameters we have not explored (Section 9.3) could increase the luminosity of these lines, our results show that coronal line emission should not be common and therefore should not used as a stand-alone metric for identifying dwarf AGN. In contrast, probing local dwarfs in the mid-IR with JWST MIRI provides the opportunity to find accreting IMBHs at low stellar masses and metallicities that evaded detection in the optical and UV (e.g., Mingozzi et al. 2025).

### 9.3. *Model Limitations*

Our model grids presented here, freely available to the community, make several assumptions that we discuss in this section. In regards to the SED of the IMBH, we derived three $M_{\rm BH} - Z$ scaling based on the observed MZR and predicted $M_{\rm BH} - M_\star$ relationship in cosmological simulations (Section 2). Undoubtedly, significant scatter exists in both scaling relations, which introduces uncertainties into our $M_{\rm BH} - Z$ correlations. While accounting for this scatter merits investigation in a future paper, this paper answers the more fundamental question: *is it necessary to account for black hole properties co-evolving with their host galaxies?* (i.e., low mass IMBHs in metal-poor gas and high mass IMBHs in more metal-rich gas)? The answer to this question is *yes*, as we detailed in Section 4, especially in extremely metal-poor galaxies. This motivates the need for greater constraints on the MZR and $M_{\rm BH} - M_\star$ at all redshifts.

Direct observational evidence for the shape of the IMBH SED is scant, and those that exist are likely highly biased (e.g., HLX-1; Titarchuk & Seifina 2016a). We selected the qsosed model given its self-consistent accretion physics, but limited our analysis to non-spinning BHs with a constant accretion rate of 0.1 $\dot{m}_{\rm edd}$. While the spin has a minor effect on the shape of the SED Kubota & Done 2018, the accretion rate can vary over several orders of magnitude and impact the predicted emission line spectrum Cann et al. 2018. A rigorous analysis with variable accretion rate is beyond the scope of this work. However, we ran a subset of models for $\dot{m} = 0.022$ and $\dot{m} = 0.5$ to assess the effect on one of the most successful mid-IR diagnostic diagrams for separating ULX and IMBH excitation. Figure 23 shows the [O IV]/[S III] vs. [S IV]/[Ar II] diagram from Figure 22 still separates active IMBHs from ULXs and stars in the case where $f_{\rm AGN} \gtrsim 0.04$ and $\dot{m} = 0.5$. In case where $\dot{m} = 0.5$, the grids mainly lie in the ULX/AGN region, but would reflect a small percentage of dwarf AGN since they are predicted to have relatively low accretion rates (Bellovary et al. 2019; Beckmann et al. 2023; Bhowmick et al. 2024).

Regarding the XRB SED, we focused strictly on ULXs with stellar mass BHs since these are most likely to cause signatures that mimic IMBH excitation. Like IMBHs, ULXs likely possess a range of accretion rates, geometries, etc., that should be explored further. We did not include low-mass X-ray binaries and Be-XRBs, which occur on different timescales than HMXBs/ULXs (>25 Myr) and might be a non-negligible population (Liu et al. 2024) to the overall ionizing SED.

It is possible that off-nuclear IMBHs can power some ULXs rather than neutron stars or stellar-mass black holes (Titarchuk & Seifina 2016a,b). However, for several reasons, most ULXs are likely to contain stellar-mass compact objects. ULXs are more likely to be found in late-type galaxies with active star formation (Bernadich et al. 2022); the unique spectral states of most ULXs require accretion largely incompatible with IMBHs (Gladstone & Roberts 2009); and simulations of IMBHs with companions show that they occur far too infrequently to reproduce the number of observed ULXs (King et al. 2008; Madhusudhan et al. 2006).

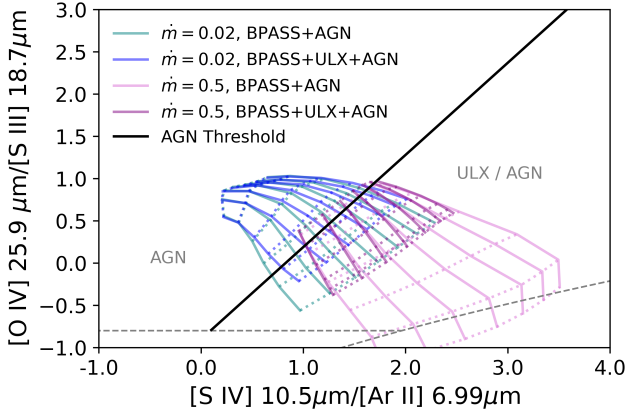

Figure 23. A subset of our models with low and high AGN accretion rates. The [O IV]/[S III] vs. [S IV]/[Ar II] diagram remains valuable for separating excitation from active IMBHs and stars.

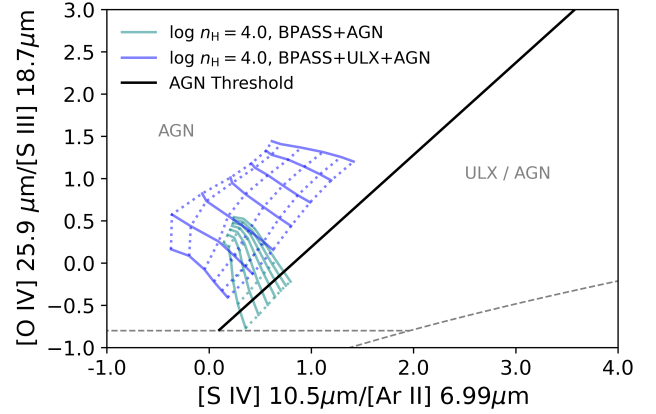

Figure 24. A subset of our models with a higher gas density of $\log n_H = 4.0$ compared our standard model that uses $\log n_H = 2.0$. The [O IV]/[S III] vs. [S IV]/[Ar II] diagram shows almost all of the AGN models residing in the AGN region.

In terms of the gas cloud properties, we limited our analysis to a single hydrogen density ($\log n_H = 2.0$) at the face of the cloud. In reality, many gas clouds with various densities exist within our modeled environments. The rationale for keeping the hydrogen density fixed comes from limiting our analysis to parameters with the most significant effect on the predicted emission line spectrum. In contrast, the gas density tends to introduce more secondary effects (Kewley et al. 2013; Lebouteiller et al. 2025), but order of magnitude variations exist (Topping et al. 2025). As such, we ran a subset of models with a much higher gas density of $\log n_H = 4.0$ as shown in Figure 24. This results in essentially all the AGN models being properly classified as AGN, adding to the robustness of the diagram for identifying active IMBHs.

We selected an $n_e/n_H = 0.01$ boundary condition for our models to focus on ionized gas emission. Including matter-bounded and ionization-bounded clouds would affect X-ray absorption and neutral gas tracers that could contain signatures of AGN excitation (Ramambason et al. 2022; Varese et al. 2024). Similarly, we assumed a constant pressure equation of state, which has been shown to hold in extragalactic HII regions (Clark et al. 2025), although in principle variations could exist. While emission lines formed in the H II region would be insensitive to changes in the density profile, those formed near the ionization front (e.g., [O I], [Ar II]) could be affected when the hard ionizing continuum from low-mass IMBHs creates an elongated partially ionized zone. Next-generation observatories like PRIMA (far-IR) and AXIS (X-ray) would benefit from predictions that would result from varying the final depth of the simulations.

The assumed geometry of our simulations is a spherically symmetric cloud where ionizing spectra from stars and the IMBH are coincident on that cloud. The way that ionizing spectra are mixed (coincident or non-coincident) can profoundly affect the predicted emission line spectrum (Richardson et al. 2022). However, the mid-IR lines, which are the most successful IMBH diagnostics in many regards, are largely insensitive to these effects. Additionally, modeling a statistical sample of clouds exposed to different continua would change our predictions (Lebouteiller et al. 2025), although this analysis can always be performed ad hoc with the existing model suite.

Finally, we stress that individual emission line diagnostics we have presented that can best identify and characterize IMBHs may trace the multiple physical parameters (e.g., $M_{BH}$, and $f_{AGN}$) and also trace the parameters we assume to be fixed (e.g., accretion rate). The solution is likely two-fold: (1) for a given physical parameter, include as many of the best emission line diagnostics as possible, which will each characterize the parameter in slightly different ways, and then use a probabilistic approach to break parameter degeneracies; (2) use different IMBH model grids that emphasize varying physical parameters we assumed to be fixed (e.g., neutral gas fraction) and then evaluate any confounding variables. A similar approach can be taken to constrain the metallicity, for example, by using multiple emission lines that originate from elements subject to secondary nucleosynthesis (e.g., nitrogen, carbon; Nicholls et al. 2017).

## 10. CONCLUSIONS

- For the first time, we have developed a cohesive framework for simultaneously modeling ULX and AGN in metal-poor, highly star-forming dwarf galaxies while accounting for multiple black hole seeding channels.

- Due to ULX contributions, we find that broadband X-ray observations (2-10 keV) are unlikely to identify active IMBHs in typical dwarf galaxies, which are highly star-forming (Section 6.2).

- UV AGN emission line diagnostics are strongly affected by ULX contributions and relatively insensitive to $M_{BH}$ and BH seeding mechanisms. However, He II $\lambda 4686$ and [O I] $\lambda 6300$ in optical, and several more mid-IR lines can identify dwarf AGN and characterize properties of the IMBH across a wide range of physical conditions even in the presence of ULXs. Other lines formed from S II or Ne V can also provide valuable constraints but have more limitations than other emission lines. A summary of the most useful line diagnostics for dwarf AGN is provided in Table 4, and the success rates for identifying dwarf AGN with common diagnostics diagrams are provided in Tables 1-3.

- We have redefined the demarcations of many diagnostic diagrams spanning the UV-IR to reflect the maximum possible ULX contribution and the approximate region where the ionizing continuum of the active IMBH dominates over ULX (Sections 8.1-8.3).

- Emission line ratios and diagnostic diagrams in the mid-IR are particularly robust in separating contributions of stars, ULXs, shocks, and AGN in dwarf galaxies. However, we caution against using diagnostic diagrams as the end-all, be-all for identifying AGN as many UV and optical diagrams fail in the metal-poor, highly star-forming (low $f_{AGN}$) regime, which can result in misclassifying excitation mechanisms or drawing broad conclusions from potentially biased samples (e.g., see Section 9.2)

- For subsequent analysis of dwarf galaxy spectra in the JWST era, we advocate for using a Bayesian approach to simultaneously use all the measured emission lines to identify the presence of an IMBH, and the *degree* to which an IMBH is active, rather than using the binary star-forming or AGN galaxy classifications.

## ACKNOWLEDGMENTS

CR gratefully acknowledges the support of the Elon University Japheth E. Rawls Professorship, Elon University's FR&D committee and the Advanced Cyberinfrastructure Coordination Ecosystem: Services & Support (ACCESS), which is supported by the National Science Foundation. This work used the ACCESS resource Expanse at the San Diego Supercomputing Center through allocation AST140040. JW acknowledges the Lumen Prize award at Elon University. JL acknowledges the support of the North Carolina Space Grant Undergraduate Research Scholarship. VL and CR acknowledge support from the Transatlantic Research Partnership. LR gratefully acknowledges funding from the DFG through an Emmy Noether Research Group (grant number CH2137/1-1). JMB is supported by NSF grant AST-2107764. We thank Marta Volonteri for her helpful discussions about this work.

*Software:* Cloudy (Ferland et al. 2017) .

## REFERENCES

Abel, N. P., Dudley, C., Fischer, J., Satyapal, S., & van Hoof, P. A. M. 2009, ApJ, 701, 1147, doi: 10.1088/0004-637X/701/2/1147

Abel, N. P., van Hoof, P. A. M., Shaw, G., Ferland , G. J., & Elwert, T. 2008, ApJ, 686, 1125, doi: 10.1086/591505

Alarie, A., & Morisset, C. 2019, RMxAA, 55, 377, doi: 10.22201/ia.01851101p.2019.55.02.21

Andrews, B. H., & Martini, P. 2013, ApJ, 765, 140, doi: 10.1088/0004-637X/765/2/140

Backhaus, B. E., Trump, J. R., Cleri, N. J., et al. 2022, ApJ, 926, 161, doi: 10.3847/1538-4357/ac3919

Backhaus, B. E., Cleri, N. J., Trump, J. R., et al. 2025, arXiv e-prints, arXiv:2502.03519. https://arxiv.org/abs/2502.03519

Baldwin, J. A., Ferland, G. J., Martin, P. G., et al. 1991, ApJ, 374, 580, doi: 10.1086/170146

Baldwin, J. A., Phillips, M. M., & Terlevich, R. 1981, PASP, 93, 5, doi: 10.1086/130766

Beckmann, R. S., Dubois, Y., Volonteri, M., et al. 2023, MNRAS, 523, 5610, doi: 10.1093/mnras/stad1544

Bellovary, J. M., Cleary, C. E., Munshi, F., et al. 2019, MNRAS, 482, 2913, doi: 10.1093/mnras/sty2842

Berg, D. A., Erb, D. K., Henry, R. B. C., Skillman, E. D., & McQuinn, K. B. W. 2019, ApJ, 874, 93, doi: 10.3847/1538-4357/ab020a

Berg, D. A., Skillman, E. D., Marble, A. R., et al. 2012, ApJ, 754, 98, doi: 10.1088/0004-637X/754/2/98

Berghea, C. T., Dudik, R. P., Weaver, K. A., & Kallman, T. R. 2010, ApJ, 708, 354, doi: 10.1088/0004-637X/708/1/354

Bernadich, M. C. i., Schwope, A. D., Kovlakas, K., Zezas, A., & Traulsen, I. 2022, A&A, 659, A188, doi: 10.1051/0004-6361/202141560

Bhat, H. K., Chakravorty, S., Sengupta, D., et al. 2020, MNRAS, 497, 2992, doi: 10.1093/mnras/staa2002

Bhowmick, A. K., Blecha, L., Torrey, P., et al. 2024, arXiv e-prints, arXiv:2402.03626, doi: 10.48550/arXiv.2402.03626

Brorby, M., Kaaret, P., Prestwich, A., & Mirabel, I. F. 2016, MNRAS, 457, 4081, doi: 10.1093/mnras/stw284

Cann, J. M., Satyapal, S., Abel, N. P., et al. 2019, ApJL, 870, L2, doi: 10.3847/2041-8213/aaf88d

—. 2018, ApJ, 861, 142, doi: 10.3847/1538-4357/aac64a

Chisholm, J., Berg, D. A., Endsley, R., et al. 2024, arXiv e-prints, arXiv:2402.18643, doi: 10.48550/arXiv.2402.18643

Cho, H., & Woo, J.-H. 2024, arXiv e-prints, arXiv:2405.09441. https://arxiv.org/abs/2405.09441

Clark, I. Y., Sandstrom, K., Wolfire, M., et al. 2025, arXiv e-prints, arXiv:2504.06247, doi: 10.48550/arXiv.2504.06247

Cleri, N. J., Olivier, G. M., Hutchison, T. A., et al. 2023, ApJ, 953, 10, doi: 10.3847/1538-4357/acde55

Curti, M., Mannucci, F., Cresci, G., & Maiolino, R. 2020, MNRAS, 491, 944, doi: 10.1093/mnras/stz2910

D'Agostino, J. J., Kewley, L. J., Groves, B., et al. 2019, ApJ, 878, 2, doi: 10.3847/1538-4357/ab1d5e

Davé, R., Anglés-Alcázar, D., Narayanan, D., et al. 2019, MNRAS, 486, 2827, doi: 10.1093/mnras/stz937

Doan, S., Satyapal, S., Matzko, W., et al. 2024, arXiv e-prints, arXiv:2408.04774, doi: 10.48550/arXiv.2408.04774

Domínguez, A., Siana, B., Brooks, A. M., et al. 2015, MNRAS, 451, 839, doi: 10.1093/mnras/stv1001

Dors, O. L., Cardaci, M. V., Hägele, G. F., et al. 2024, MNRAS, 527, 8193, doi: 10.1093/mnras/stad3667

Dray, L. M., & Tout, C. A. 2007, MNRAS, 376, 61, doi: 10.1111/j.1365-2966.2007.11431.x

Einig, L., Palud, P., Roueff, A., et al. 2024, A&A, 691, A109, doi: 10.1051/0004-6361/202451588

Eldridge, J. J., Stanway, E. R., Xiao, L., et al. 2017, PASA, 34, e058, doi: 10.1017/pasa.2017.51

Feltre, A., Charlot, S., & Gutkin, J. 2016, MNRAS, 456, 3354, doi: 10.1093/mnras/stv2794

Feltre, A., Gruppioni, C., Marchetti, L., et al. 2023, A&A, 675, A74, doi: 10.1051/0004-6361/202245516

Ferland, G. J., Chatzikos, M., Guzmán, F., et al. 2017, RMxAA, 53, 385. https://arxiv.org/abs/1705.10877

Fragos, T., Lehmer, B., Tremmel, M., et al. 2013, ApJ, 764, 41, doi: 10.1088/0004-637X/764/1/41

Garofali, K., Basu-Zych, A. R., Johnson, B. D., et al. 2024, ApJ, 960, 13, doi: 10.3847/1538-4357/ad0a6a

Geha, M., Blanton, M. R., Yan, R., & Tinker, J. L. 2012, ApJ, 757, 85, doi: 10.1088/0004-637X/757/1/85

Gladstone, J. C., & Roberts, T. P. 2009, MNRAS, 397, 124, doi: 10.1111/j.1365-2966.2009.14937.x

Götberg, Y., de Mink, S. E., Groh, J. H., et al. 2018, A&A, 615, A78, doi: 10.1051/0004-6361/201732274

Greene, J. E., Strader, J., & Ho, L. C. 2020, ARA&A, 58, 257, doi: 10.1146/annurev-astro-032620-021835

Greene, J. E., Labbe, I., Goulding, A. D., et al. 2024, ApJ, 964, 39, doi: 10.3847/1538-4357/ad1e5f

Gunasekera, C. M., Ji, X., Chatzikos, M., Yan, R., & Ferland, G. 2022, MNRAS, 512, 2310, doi: 10.1093/mnras/stac022

Gúrpide, A., Castro Segura, N., Soria, R., & Middleton, M. 2024, arXiv e-prints, arXiv:2405.13714, doi: 10.48550/arXiv.2405.13714

Gutkin, J., Charlot, S., & Bruzual, G. 2016, MNRAS, 462, 1757, doi: 10.1093/mnras/stw1716

Haidar, H., Habouzit, M., Volonteri, M., et al. 2022, MNRAS, 514, 4912, doi: 10.1093/mnras/stac1659

Harikane, Y., Zhang, Y., Nakajima, K., et al. 2023, ApJ, 959, 39, doi: 10.3847/1538-4357/ad029e

Henry, R. B. C., Edmunds, M. G., & Köppen, J. 2000, ApJ, 541, 660, doi: 10.1086/309471

Indahl, B., Zeimann, G., Hill, G. J., et al. 2021, ApJ, 916, 11, doi: 10.3847/1538-4357/ac01ed

Jaskot, A. E., & Ravindranath, S. 2016, ApJ, 833, 136, doi: 10.3847/1538-4357/833/2/136

Jenkins, E. B. 2009, ApJ, 700, 1299, doi: 10.1088/0004-637X/700/2/1299

Jenkins, E. B., & Wallerstein, G. 2017, ApJ, 838, 85, doi: 10.3847/1538-4357/aa64d4

Kewley, L. J., Dopita, M. A., Leitherer, C., et al. 2013, ApJ, 774, 100, doi: 10.1088/0004-637X/774/2/100

Kewley, L. J., Dopita, M. A., Sutherland, R. S., Heisler, C. A., & Trevena, J. 2001, ApJ, 556, 121, doi: 10.1086/321545

Kewley, L. J., & Ellison, S. L. 2008, ApJ, 681, 1183, doi: 10.1086/587500

Kewley, L. J., Groves, B., Kauffmann, G., & Heckman, T. 2006, MNRAS, 372, 961, doi: 10.1111/j.1365-2966.2006.10859.x

King, A., Lasota, J.-P., & Middleton, M. 2023, NewAR, 96, 101672, doi: 10.1016/j.newar.2022.101672

King, A. R., Davies, M. B., Ward, M. J., Fabbiano, G., & Elvis, M. 2001, ApJL, 552, L109, doi: 10.1086/320343

King, A. R., Pringle, J. E., & Hofmann, J. A. 2008, MNRAS, 385, 1621, doi: 10.1111/j.1365-2966.2008.12943.x

Kobayashi, C., Tominaga, N., & Nomoto, K. 2011, ApJL, 730, L14, doi: 10.1088/2041-8205/730/2/L14

Kosec, P., Pinto, C., Fabian, A. C., & Walton, D. J. 2018, MNRAS, 473, 5680, doi: 10.1093/mnras/stx2695

Kroupa, P. 2002, Science, 295, 82, doi: 10.1126/science.1067524

Kubota, A., & Done, C. 2018, MNRAS, 480, 1247, doi: 10.1093/mnras/sty1890

Lamareille, F. 2010, A&A, 509, A53, doi: 10.1051/0004-6361/200913168

Langeroodi, D., Hjorth, J., Chen, W., et al. 2023, ApJ, 957, 39, doi: 10.3847/1538-4357/acdbc1

Lebouteiller, V., & Ramambason, L. 2022, A&A, 667, A34, doi: 10.1051/0004-6361/202243865

Lebouteiller, V., Richardson, C. T., Polimera, M. S., et al. 2025, A&A, 695, A31, doi: 10.1051/0004-6361/202452103

Lecroq, M., Charlot, S., Bressan, A., et al. 2024, MNRAS, 527, 9480, doi: 10.1093/mnras/stad3838

Lee, J. C., Kennicutt, R. C., Funes, José G., S. J., Sakai, S., & Akiyama, S. 2007, ApJL, 671, L113, doi: 10.1086/526341

Lehmer, B. D., Basu-Zych, A. R., Mineo, S., et al. 2016, ApJ, 825, 7, doi: 10.3847/0004-637X/825/1/7

Lehmer, B. D., Eufrasio, R. T., Basu-Zych, A., et al. 2021, ApJ, 907, 17, doi: 10.3847/1538-4357/abcec1

Leitherer, C., Schaerer, D., Goldader, J. D., et al. 1999, ApJS, 123, 3, doi: 10.1086/313233

Li, S.-L., Grasha, K., Krumholz, M. R., et al. 2024, MNRAS, doi: 10.1093/mnras/stae869

Linden, T., Kalogera, V., Sepinsky, J. F., et al. 2010, ApJ, 725, 1984, doi: 10.1088/0004-637X/725/2/1984

Liu, B., Sartorio, N. S., Izzard, R. G., & Fialkov, A. 2024, MNRAS, 527, 5023, doi: 10.1093/mnras/stad3475

Lupi, A., Trinca, A., Volonteri, M., Dotti, M., & Mazzucchelli, C. 2024, arXiv e-prints, arXiv:2406.17847, doi: 10.48550/arXiv.2406.17847

Mac Low, M.-M., & Ferrara, A. 1999, ApJ, 513, 142, doi: 10.1086/306832

Madhusudhan, N., Justham, S., Nelson, L., et al. 2006, ApJ, 640, 918, doi: 10.1086/500238

Maiolino, R., & Mannucci, F. 2019, A&A Rv, 27, 3, doi: 10.1007/s00159-018-0112-2

Maiolino, R., Scholtz, J., Curtis-Lake, E., et al. 2023, arXiv e-prints, arXiv:2308.01230, doi: 10.48550/arXiv.2308.01230

Mannucci, F., Cresci, G., Maiolino, R., Marconi, A., & Gnerucci, A. 2010, MNRAS, 408, 2115, doi: 10.1111/j.1365-2966.2010.17291.x

Mapelli, M., Ripamonti, E., Zampieri, L., Colpi, M., & Bressan, A. 2010, MNRAS, 408, 234, doi: 10.1111/j.1365-2966.2010.17048.x

Mazzolari, G., Übler, H., Maiolino, R., et al. 2024, arXiv e-prints, arXiv:2404.10811, doi: 10.48550/arXiv.2404.10811

McAlpine, S., Helly, J. C., Schaller, M., et al. 2016, Astronomy and Computing, 15, 72, doi: 10.1016/j.ascom.2016.02.004

Mezcua, M. 2017, International Journal of Modern Physics D, 26, 1730021, doi: 10.1142/S021827181730021X

Mezcua, M., & Domínguez Sánchez, H. 2020, ApJL, 898, L30, doi: 10.3847/2041-8213/aba199

—. 2024, MNRAS, 528, 5252, doi: 10.1093/mnras/stae292

Mezcua, M., Pacucci, F., Suh, H., Siudek, M., & Natarajan, P. 2024, arXiv e-prints, arXiv:2404.05793, doi: 10.48550/arXiv.2404.05793

Miller, B. P., Gallo, E., Greene, J. E., et al. 2015, ApJ, 799, 98, doi: 10.1088/0004-637X/799/1/98

Mingozzi, M., James, B. L., Berg, D. A., et al. 2024, ApJ, 962, 95, doi: 10.3847/1538-4357/ad1033

Mitsuda, K., Inoue, H., Koyama, K., et al. 1984, PASJ, 36, 741

Molina, M., Reines, A. E., Latimer, L. J., Baldassare, V., & Salehirad, S. 2021, ApJ, 922, 155, doi: 10.3847/1538-4357/ac1ffa

Morisset, C., Charlot, S., Sánchez, S. F., et al. 2025, arXiv e-prints, arXiv:2501.05424, doi: 10.48550/arXiv.2501.05424

Nakajima, K., Schaerer, D., Le Fèvre, O., et al. 2018, A&A, 612, A94, doi: 10.1051/0004-6361/201731935

Negus, J., Comerford, J. M., Sánchez, F. M., et al. 2023, ApJ, 945, 127, doi: 10.3847/1538-4357/acb772

Nguyen, D. D., Seth, A. C., Neumayer, N., et al. 2018, ApJ, 858, 118, doi: 10.3847/1538-4357/aabe28

Nicholls, D. C., Sutherland, R. S., Dopita, M. A., Kewley, L. J., & Groves, B. A. 2017, MNRAS, 466, 4403, doi: 10.1093/mnras/stw3235

Panda, S., Czerny, B., Done, C., & Kubota, A. 2019, ApJ, 875, 133, doi: 10.3847/1538-4357/ab11cb

Peterson, B. M. 1997, An Introduction to Active Galactic Nuclei

Pharo, J., Guo, Y., Calvo, G. B., et al. 2023, ApJ, 959, 48, doi: 10.3847/1538-4357/ad0134

Pinto, C., Middleton, M. J., & Fabian, A. C. 2016, Nature, 533, 64, doi: 10.1038/nature17417

Polimera, M. S., Kannappan, S. J., Richardson, C. T., et al. 2022, ApJ, 931, 44, doi: 10.3847/1538-4357/ac6595

Ramambason, L., Lebouteiller, V., Bik, A., et al. 2022, A&A, 667, A35, doi: 10.1051/0004-6361/202243866

Ramambason, L., Lebouteiller, V., Madden, S. C., et al. 2024, A&A, 681, A14, doi: 10.1051/0004-6361/202347280

Regan, J., & Volonteri, M. 2024, arXiv e-prints, arXiv:2405.17975, doi: 10.48550/arXiv.2405.17975

Reines, A. E., Condon, J. J., Darling, J., & Greene, J. E. 2020, ApJ, 888, 36, doi: 10.3847/1538-4357/ab4999

Reines, A. E., & Volonteri, M. 2015, ApJ, 813, 82, doi: 10.1088/0004-637X/813/2/82

Rémy-Ruyer, A., Madden, S. C., Galliano, F., et al. 2014, A&A, 563, A31, doi: 10.1051/0004-6361/201322803

Reyero Serantes, S., Oskinova, L., Hamann, W. R., et al. 2024, A&A, 690, A347, doi: 10.1051/0004-6361/202451324

Ricarte, A., & Natarajan, P. 2018, MNRAS, 481, 3278, doi: 10.1093/mnras/sty2448

Richardson, C. T., Simpson, C., Polimera, M. S., et al. 2022, ApJ, 927, 165, doi: 10.3847/1538-4357/ac510c

Salehirad, S., Reines, A. E., & Molina, M. 2022, ApJ, 937, 7, doi: 10.3847/1538-4357/ac8876

Sarkar, A., Ferland, G. J., Chatzikos, M., et al. 2021, ApJ, 907, 12, doi: 10.3847/1538-4357/abcaa6

Sartori, L. F., Schawinski, K., Treister, E., et al. 2015, MNRAS, 454, 3722, doi: 10.1093/mnras/stv2238

Satyapal, S., Kamal, L., Cann, J. M., Secrest, N. J., & Abel, N. P. 2021, ApJ, 906, 35, doi: 10.3847/1538-4357/abbfaf

Schneider, F. R. N., Podsiadlowski, P., Langer, N., Castro, N., & Fossati, L. 2016, MNRAS, 457, 2355, doi: 10.1093/mnras/stw148

Senchyna, P., Stark, D. P., Mirocha, J., et al. 2020, MNRAS, 494, 941, doi: 10.1093/mnras/staa586

Shirazi, M., & Brinchmann, J. 2012, MNRAS, 421, 1043, doi: 10.1111/j.1365-2966.2012.20439.x

Sijacki, D., Vogelsberger, M., Genel, S., et al. 2015, MNRAS, 452, 575, doi: 10.1093/mnras/stv1340

Simmonds, C., Schaerer, D., & Verhamme, A. 2021, A&A, 656, A127, doi: 10.1051/0004-6361/202141856

Smith, N. 2014, ARA&A, 52, 487, doi: 10.1146/annurev-astro-081913-040025

Suh, H., Civano, F., Trakhtenbrot, B., et al. 2020, ApJ, 889, 32, doi: 10.3847/1538-4357/ab5f5f

Thomas, A. D., Dopita, M. A., Kewley, L. J., et al. 2018, ApJ, 856, 89, doi: 10.3847/1538-4357/aab3db

Ting, Y.-S., & Ji, A. P. 2024, arXiv e-prints. https://arxiv.org/abs/2408.06807

Titarchuk, L., & Seifina, E. 2016a, A&A, 595, A101, doi: 10.1051/0004-6361/201527840

—. 2016b, A&A, 585, A94, doi: 10.1051/0004-6361/201526122

Topping, M. W., Sanders, R. L., Shapley, A. E., et al. 2025, arXiv e-prints, arXiv:2502.08712. https://arxiv.org/abs/2502.08712

Torrey, P., Vogelsberger, M., Marinacci, F., et al. 2019, MNRAS, 484, 5587, doi: 10.1093/mnras/stz243

Trani, A. A., Mapelli, M., & Bressan, A. 2014, MNRAS, 445, 1967, doi: 10.1093/mnras/stu1898

Tremonti, C. A., Heckman, T. M., Kauffmann, G., et al. 2004, ApJ, 613, 898, doi: 10.1086/423264

Varese, M., Lebouteiller, V., Ramambason, L., et al. 2024, arXiv e-prints, arXiv:2411.03912, doi: 10.48550/arXiv.2411.03912

Veilleux, S., & Osterbrock, D. E. 1987, ApJS, 63, 295, doi: 10.1086/191166

Volonteri, M., Lodato, G., & Natarajan, P. 2008, MNRAS, 383, 1079, doi: 10.1111/j.1365-2966.2007.12589.x

Weaver, K. A., Meléndez, M., Mushotzky, R. F., et al. 2010, ApJ, 716, 1151, doi: 10.1088/0004-637X/716/2/1151

West, L., Garofali, K., Lehmer, B. D., et al. 2023, ApJ, 952, 22, doi: 10.3847/1538-4357/acd9aa

Wiktorowicz, G., Sobolewska, M., Lasota, J.-P., & Belczynski, K. 2017, ApJ, 846, 17, doi: 10.3847/1538-4357/aa821d

Xiao, L., Stanway, E. R., & Eldridge, J. J. 2018, MNRAS, 477, 904, doi: 10.1093/mnras/sty646